\documentclass[referee]{raa}
\usepackage{graphicx,times}
\usepackage{natbib}
\usepackage{amssymb,amsmath}
\bibpunct{(}{)}{;}{a}{}{,}

\usepackage[a4paper=true,dvipdfm=true,pagebackref=true]{hyperref}
\hypersetup{pdftitle = The title of my PDF, pdfauthor = My name, pdfsubject= The subject, pdfkeywords = keyword1 keyword2 keyword3}
\hypersetup{colorlinks = true, linkcolor = green, anchorcolor = red, citecolor = blue, filecolor = red, pagecolor = red, urlcolor = red}
\usepackage{multirow}

\begin{document}

   \title{A preliminary study of photometric redshifts based on the Wide Field Survey Telescope}

 \volnopage{ {\bf 20XX} Vol.\ {\bf X} No. {\bf XX}, 000--000}
   \setcounter{page}{1}

   \author{Yu Liu\inst{1,2}, Xiao-zhi Lin\inst{1,2,3}, Yong-quan Xue\inst{1,2} and Huynh Anh N. Le\inst{1,2}
   }

   \institute{ CAS Key Laboratory for Research in Galaxies and Cosmology, Department of Astronomy, University of Science and Technology of China, Hefei 230026, China; xzlin@ustc.edu.cn, xuey@ustc.edu.cn\\
        \and
             School of Astronomy and Space Sciences, University of Science and Technology of China, Hefei 230026, China\\
	\and
Xiao-zhi Lin and Yu Liu contributed equally to this work.\\
\vs \no
   {\small Received 20XX Month Day; accepted 20XX Month Day}
}

\abstract{The Wide Field Survey Telescope (WFST) is a
dedicated time-domain multi-band ($u$, $g$, $r$, $i$, and $z$) photometric survey facility under construction.
In this paper, we present a preliminary study that assesses the quality of photometric redshifts based on WFST by utilizing mock observations derived with the galaxy catalog in the COSMOS/UltraVISTA field. We apply the template fitting technique to estimate photometric redshifts by using the ZEBRA photometric-redshift code and adopting a modified set of adaptive templates. We evaluate the bias (median relative offset between the output photometric redshifts and input redshifts), normalized median absolute deviation ($\sigma_{\rm NMAD}$) and outlier fraction ($f_{\rm outlier}$) of photometric redshifts in two typical WFST observational cases, the single 30-second exposure observations (hereafter shallow mode) and co-added 50-minute exposure observations (hereafter deep mode).
We find bias$\la0.006$, $\sigma_{\rm NMAD}\la0.03$, and $f_{\rm outlier}\la5\%$ in the shallow mode and
bias$\approx 0.005$, $\sigma_{\rm NMAD}\approx 0.06$, and $f_{\rm outlier}\approx 17\%$--$27\%$ in the deep mode, respectively, under various lunar phases.
Combining the WFST mock observational data with that from the upcoming CSST and Euclid surveys, we demonstrate that the $z_{\rm phot}$ results can be significantly improved, with
$f_{\rm outlier}\approx 1\%$ and $\sigma_{\rm NMAD}\approx 0.02$.
}

   \authorrunning{Y. Liu et al. }            
   \titlerunning{Photometric redshifts based on WFST}  
   \maketitle

%
\section{Introduction}           
\label{sect:intro}

\hspace{5mm}The development of modern astronomy has given rise to an increasing demand for powerful multi-band photometric sky surveys.
Such surveys, e.g., the Sloan Digital Sky Survey \citep[SDSS; e.g.,][]{Brescia_2014,Albareti_2017,Zhao_2021}, Dark Energy Survey \citep[DES; e.g.,][]{Drinkwater_2010,DESCollaboration_2016,Ivezic_2019}, and Hyper Suprime-Cam Subaru Strategic Program Survey \citep[HSC-SSP; e.g.,][]{Aihara_2018,Hikage_2019}, with well-designed equipments, reasonable observational strategies, and fruitful scientific results in stellar physics, galaxy physics, and cosmology, have demonstrated their strong impacts on modern astronomy.

The Wide Field Survey Telescope (WFST) is a
dedicated time-domain multi-band ($u$, $g$, $r$, $i$, and $z$) photometric survey facility under construction jointly by the University of Science and Technology of China and Purple Mountain Observatory, which is expected to start commissioning observations around August 2023.
WFST has a 2.5-meter primary mirror, an active optical system, and a 0.73-Gigapixel mosaic CCD camera on the main focus plane;
moreover, WFST is located near the summit of the Saishiteng Mountain in the Lenghu area that is a world-class observational site \citep{Deng_2021},
thereby achieving high-quality imaging over a field of view of 6.5 deg$^2$.
The main science goals of WFST surveys are time-domain sciences including supernovae, tidal disruption events, multi-messenger events, and active galactic nuclei (AGNs), asteroids and the solar system, the Milky Way and its satellite dwarf galaxies, and galaxy formation and cosmology \citep{WFST_science}.

Robust determination of cosmological redshifts is one of the most crucial factors in fulfilling the above WFST science goals.
However, high-precision galaxy redshift measurements require spectroscopic observations for each source (i.e., obtaining spectroscopic redshifts, $z_{\rm spec}$).
This task is not only expensive but also time consuming.
Alternatively, there is another way to measure redshifts using photometric surveys (i.e., obtaining photometric redshifts, $z_{\rm phot}$), which is much more efficient than spectroscopic observations.
This method, although not as precise as the $z_{\rm spec}$ measurement, has demonstrated its extensive use in the $z_{\rm phot}$ determination of an overall huge amount of survey targets at one time \citep[e.g.,][]{Benjamin_2010,Brescia_2014,Cavuoti_2017,Sanchez_2019}.
The application of $z_{\rm phot}$ has enabled a wide range of exciting extragalactic sciences as mentioned above.

To date, a series of methods have been developed to estimate $z_{\rm phot}$.
In general, they can be divided into two main categories.
One is based on template fitting that works as follows: the observed photometry is compared to a given set of pre-assumed galaxy templates to determine the best-fit redshift corresponding to the maximum likelihood \citep[e.g.,][]{Benitez_2000,ZEBRA,EAZY,Luo_2010,Rafferty_2011,Yang_2014,Cao_2018}.
The other is the so-called training-set method, which constructs a neural network \citep[e.g.,][]{ANNZ,Blake_2007,Sanchez_2014,Pasquet_2019} and performs machine learning to obtain $z_{\rm phot}$, focusing on finding empirical relations between the redshift and galaxy properties (e.g., magnitudes and colors).
This method is usually based on a large sample of secure $z_{\rm spec}$, which are mostly available in the lower-redshift universe.
However, since the magnitude limits of all WFST bands are deeper than most of the current $z_{\rm spec}$ surveys, it is difficult to find a sample of well-measured $z_{\rm spec}$ that can be representative of the full survey sample.
Therefore, in this paper, we choose to measure $z_{\rm phot}$ of mock WFST observations based on the former technique, i.e., template fitting.

The main goal of this paper is to assess $z_{\rm phot}$ quality of the WFST photometry system preliminarily.
We utilize the COSMOS/UltraVISTA multiwavelength galaxy photometry catalog \citep{Muzzin_2013a} to produce mock WFST data.
This survey has deeper magnitude limits than WFST observations, so
it is suitable to select a subsample of galaxies from this survey, the magnitudes of which meet the WFST detection limits.
Utilizing this subsample, we generate the mock flux of each WFST filter passband based on WFST instrumental parameters with good data quality, and then estimate the corresponding observational error.
We choose to use the ZEBRA code \citep{ZEBRA} for $z_{\rm phot}$ estimation.
The main advantage of this code is that it can generate a new set of templates adaptive to the observations to minimize the mismatch between observed spectral energy distributions (SEDs) and the galaxy templates that are either from theoretical synthesis models or observed certain types of galaxy SEDs in the local universe, thereby improving $z_{\rm phot}$ quality.

This paper is organized as follows. In Section~\ref{sect:data}, we introduce the WFST photometry system, COSMOS/UltraVISTA galaxy catalog, and generation of mock WFST data;
in Section~\ref{sect:z_compute}, we introduce the process of $z_{\rm phot}$ computation;
in Section~\ref{sect:R&D}, we show $z_{\rm phot}$ results and make comparisons with other works; and
in Section~\ref{sect:sum}, we summarize our results.
All the magnitudes quoted are AB magnitudes.

\section{Data}\label{sect:data}

\subsection{Overview of the WFST photometry system}\label{subsect:WFST_overview}

\begin{figure}
   \centering
   \includegraphics[width=12.0cm, angle=0]{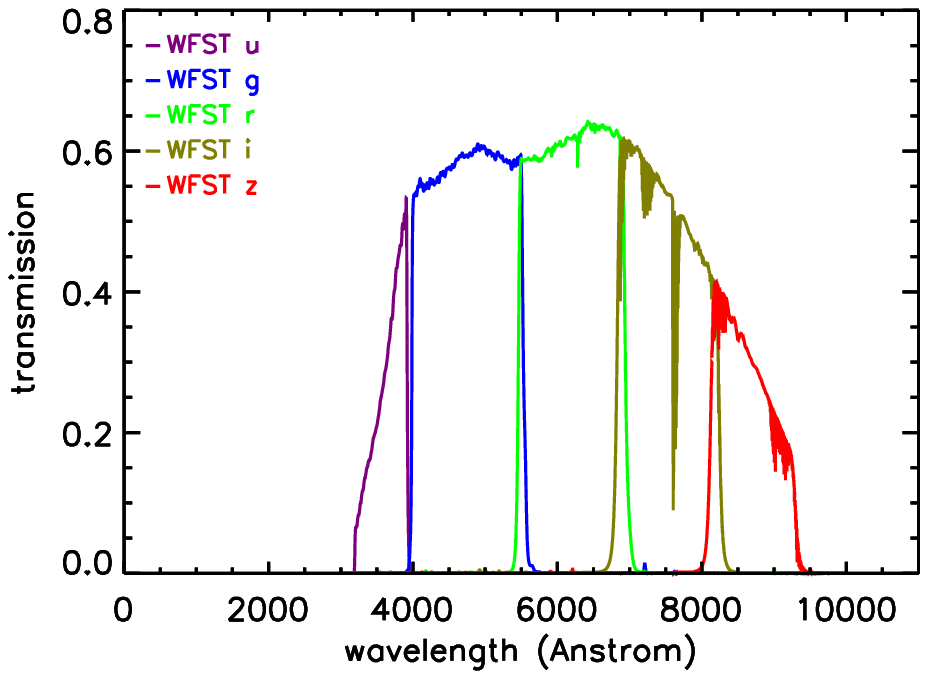}
   \caption{Total transmission curves of the WFST $ugriz$ filters (indicated by different colors), with the instrument response and atmosphere absorption and scattering taken into account.}
   \label{fig:filters}
   \end{figure}

\hspace{5mm} WFST has six filters, i.e., $u$, $g$, $r$, $i$, $z$ (see Figure~\ref{fig:filters}) and $w$, with the white-light $w$ band specifically designed for detecting asteroids in the solar system and thus being excluded from $z_{\rm phot}$ computation in this paper.
There are two planned key programs of the 6-year WFST survey: the wide-field survey (WFS) program and the deep high-cadence $u$-band survey (DHS) program.
The WFS program aims to survey a total of $\approx 8000~{\rm deg}^2$ sky area in the $u$, $g$, $r$, and $i$ bands in the northern hemisphere, with about 90 visits in each band over 6 years given a single exposure of 30 seconds for each visit; while the DHS program plans to routinely monitor a total of $\approx 720~{\rm deg}^2$ sky area in the highly sensitive $u$ band surrounding the equator every year, with a much higher observing cadence (down to hours) and being supplemented by a multi-band ancillary survey.
The $z$-band imaging is excluded in the WFS program due to its relatively low efficiency and limited contribution to time-domain sciences;
moreover, high-quality z-band imaging data will be achieved by other northern-hemisphere surveys such as Wide Imaging with Subaru HSC of the Euclid Sky (WISHES).
However, WFST will allocate some additional observational time (about 1,300 hours over 6 years) for specific purposes or particular interests, e.g., capturing time-critical targets and mapping the Galactic plane, which require intensive scanning of certain sky areas using the $z$-band imaging.

In this paper, we compute $z_{\rm phot}$ in two typical WFST observational cases, i.e., the single 30-second exposure observations (hereafter shallow mode) and co-added 50-minute exposure observations (hereafter deep mode). The deep mode can be realized by integrating all the observational time in each band mainly with the WFS program, thus achieving deeper detection limits than any existing single-telescope surveys with comparable survey areas in the northern hemisphere \citep{Lei_2023,WFST_science}.

The average night sky background brightness at the WFST site (i.e., the Saishiteng Mountain, Lenghu Town, Qinghai Province) is approximately $V=22.0~{\rm mag~arcsec^{-2}}$ when the moon is below the horizon;
under new moon conditions, the best sky level can reach $22.3~{\rm mag~arcsec^{-2}}$, which is measured in the extreme case when the bright part of the Galactic Disk is far away from the local zenith \citep{Deng_2021}.
In this paper, we set the sky background to a fixed value of $V=22.3~{\rm mag~arcsec^{-2}}$.
Under this circumstance and with no moon, the $5\sigma$ limiting magnitudes can reach depths of $ugriz=[22.31,23.42,22.95,22.43,21.50]$ in the shallow mode and $ugriz=[24.86,25.95,25.48,24.96,24.03]$ in the deep mode, respectively \citep{WFST_science}.
The modeling results of the $5\sigma$ limiting magnitudes and sky backgrounds in different lunar phases are listed in Table~\ref{tab:WFST_depth} \citep{Lei_2023}.

\begin{table*} \caption{5-$\sigma$ limiting magnitudes and sky backgrounds of WFST observations \citep[from][]{Lei_2023}}
\centering

\begin{tabular}{lccccccccc}
\hline
\hline
 Lunar Phase & Observational mode & WFST-$u$ & WFST-$g$ & WFST-$r$ & WFST-$i$ & WFST-$z$ \\
\hline
\multirow{3}{2cm}{$0~{\rm deg}$ (no moon)} & Shallow mode & 22.31 & 23.42 & 22.95 & 22.43 & 21.50 \\
~ & Deep mode & 24.86 & 25.95 & 25.48 & 24.96 & 24.03 \\
~ & Sky background & 23.27 & 22.82 & 21.80 & 20.99 & 20.05 \\
\hline
\multirow{3}{2cm}{$45~{\rm deg}$ (1/4 moon)} & Shallow mode & 22.27 & 23.30 & 22.89 & 22.40 & 21.49 \\
~ & Deep mode & 24.82 & 25.84 & 25.42 & 24.93 & 24.02 \\
~ & Sky background & 23.02 & 22.49 & 21.66 & 20.93 & 20.03 \\
\hline
\multirow{3}{2cm}{$90~{\rm deg}$ (half moon)} & Shallow mode & 22.04 & 22.86 & 22.62 & 22.26 & 21.43 \\
~ & Deep mode & 24.58 & 25.38 & 25.14 & 24.78 & 23.96 \\
~ & Sky background & 22.00 & 21.37 & 20.99 & 20.61 & 19.9 \\
\hline
\multirow{3}{2cm}{$135~{\rm deg}$ (3/4 moon)} & Shallow mode & 21.64 & 22.34 & 22.21 & 21.99 & 21.31 \\
~ & Deep mode & 24.17 & 24.85 & 24.72 & 24.51 & 23.83 \\
~ & Sky background & 20.86 & 20.21 & 20.08 & 20.01 & 19.61 \\
\hline
\multirow{3}{2cm}{$180~{\rm deg}$ (full moon)} & Shallow mode & 20.97 & 21.62 & 21.58 & 21.49 & 21.00 \\
~ & Deep mode & 23.48 & 24.12 & 24.09 & 24.01 & 23.51 \\
~ & Sky background & 19.30 & 18.73 & 18.78 & 18.97 & 18.92 \\
\hline
\end{tabular}
\label{tab:WFST_depth}
\end{table*}

\subsection{The COSMOS/UltraVISTA galaxy catalog}\label{subsect:mock}

\hspace{5mm}In this paper, we adopt the multiwavelength galaxy photometry catalog in the COSMOS/UltraVISTA field \citep{Muzzin_2013a} to produce mock WFST data,
given that it has deep optical coverage, broadband photometry, and high-quality $z_{\rm phot}$ and corresponding best-fit galaxy SEDs.

This catalog covers a sky area of 1.62 $\rm deg^2$ with point-spread function (PSF) matched photometry in 30 bands, with
the wavelength range extending from $0.15~\mu {\rm m}$ to $24~\mu {\rm m}$, including 2 ultraviolet bands (FUV and NUV) from the $GALEX$ satellite \citep{Martin_2005}, 7 broadband ($u^*$, $g^+$, $r^+$, $i^+$, $z^+$, $B_j$, $V_j$) and 12 medium-band (IA427--IA827) optical data from the Subaru and Canada-France-Hawaii Telescope \citep{Taniguchi_2007,Capak_2007}, 4 near-infrared imaging bands ($Y,J,H,K_s$) from the UltraVISTA survey \citep{McCracken_2012}, and the $3.6~\mu{\rm m}$, $4.5~\mu{\rm m}$, $5.8~\mu{\rm m}$, $8.0~\mu{\rm m}$, and $24~\mu{\rm m}$ channels from $Spitzer$'s IRAC and MIPS cameras \citep{Sanders_2007}.
The $5\sigma$ depths of the COSMOS/UltraVISTA survey in all bands are tabulated in Table~\ref{tab:COSMOS_depth}, with typical depths in optical bands being deeper than those of WFST (see Table~\ref{tab:WFST_depth}).

\begin{table*} \caption{Depths of the 30 bands in the COSMOS/UltraVISTA photometry catalog}
\centering
\begin{tabular}{lcccccccccc}
\hline
\hline
Filter & FUV & NUV & $u^*$ & $B_j$ & $g_+$ & $V_j$ & $r^+$ & $i^+$ \\
$5\sigma$ Depth & 25.2 & 25.1 & $26.4$ & $27.3$ & $27.0$ & $26.6$ & $26.8$ & $26.2$ \\
\hline
Filter & $z^+$ & IA427 & IA464 & IA484 & IA505 & IA527 & IA574 & IA624\\
$5\sigma$ Depth & $25.2$ & $25.8$ & $25.6$ & $25.9$ & $25.6$ & $25.7$ & $25.4$ & $25.7$\\
\hline
Filter & IA679 & IA709 & IA738 & IA767 & IA827 & $Y$ & $J$ & $H$\\
$5\sigma$ Depth & $25.3$ & $25.4$ & $25.4$ & $25.1$ & $25.1$ & $24.6$ & $24.4$ & $23.9$\\
\hline
Filter & $K_s$ & $3.6~\mu{\rm m}$ & $4.5~\mu {\rm m}$ & $5.8~\mu{\rm m}$ & $8.0~\mu{\rm m}$ & $24~\mu{\rm m}$\\
$5\sigma$ Depth & $23.7$ & 23.9 & 23.3 & 21.3 & 21.0 & $45~\mu{\rm Jy}$\\
\hline
\hline
\end{tabular}
\label{tab:COSMOS_depth}
\end{table*}

\begin{figure}
   \centering
   \includegraphics[width=0.8\columnwidth, angle=0]{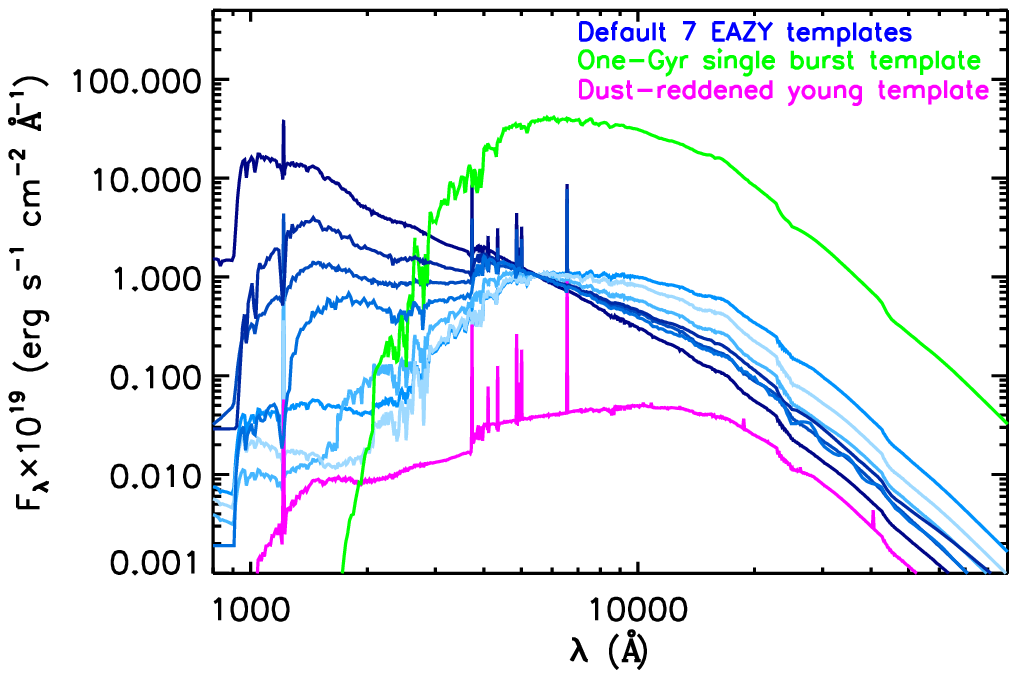}
   \caption{The set of 9 initial galaxy templates adopted by both \citet{Muzzin_2013a} and this work for $z_{\rm phot}$ derivation (the templates have been normalized appropriately for displaying purpose; see main texts for details).
}
   \label{fig:templates}
   \end{figure}

\begin{figure}
   \centering
   \includegraphics[width=14.0cm, angle=0]{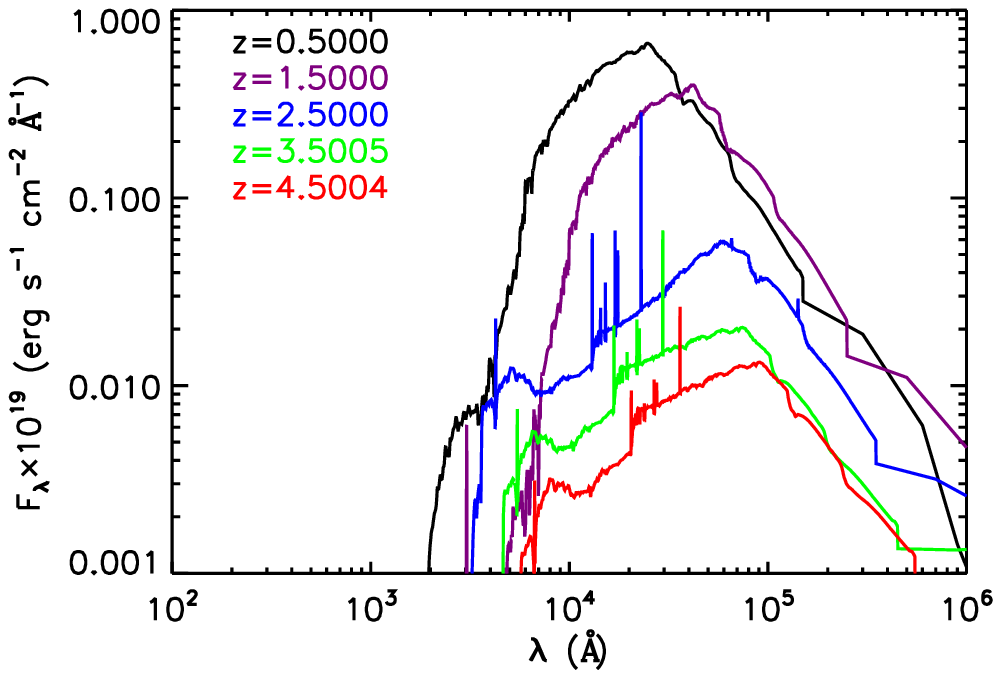}
   \caption{Some best-fit galaxy SED examples in the observed frame in the COSMOS/UltraVISTA field \citep{Muzzin_2013a}.}
   \label{fig:SEDs}
   \end{figure}

Photometric redshifts of galaxies in the COSMOS/UltraVISTA catalog are computed based on the template-fitting technique with the EAZY photometric-redshift code \citep{EAZY}.
The default 7 EAZY templates are comprised of six templates derived from the PEGASE models \citep{FR_1999} and a red galaxy template from the models of \citet{M05}.
To improve the quality of the fitting, \citet{Muzzin_2013a} added two additional galaxy templates:
one is a one-gigayear-old single-burst galaxy template generated from the \citet{BC03} model to improve the template fitting for galaxies at $z>1$ with post starburst-like features; and
the other is a slightly dust-reddened young galaxy template to improve the fitting of $UV$ bright Lyman break galaxies (LBGs) with heavy dust extinction at $1.5<z<3.5$.
EAZY fits the observed multiwavelength photometry of galaxies utilizing linear combinations of the above 9 initial templates (as shown in Figure.~\ref{fig:templates}) based on the $\chi^2$ minimization algorithm.
\citet{Muzzin_2013a} provided in their COSMOS/UltraVISTA catalog the best template combination coefficients for each of the galaxies, so that we can generate its best-fit SED.
We show some of the best-fit galaxy SED examples at their respective redshifts from the COSMOS/UltraVISTA catalog in Figure~\ref{fig:SEDs}.
Photometric redshifts derived by \citet{Muzzin_2013a} are of high quality, being consistent with $z_{\rm spec}$ from the zCOSMOS survey:
up to $z\sim1.5$, their $z_{\rm phot}$ are accurate to $\Delta z/(1+z)=0.013$, with an outlier fraction of only 1.6\%;
up to $z\sim3$, their $z_{\rm phot}$ show good agreement with $z_{\rm phot}$ from the NEWFIRM Medium Band Survey.

\subsection{Generation of mock WFST data}\label{subsect:mock_flux}

\hspace{5mm} First, the mock flux in each band for each galaxy in the given catalog can be calculated by convolving the galaxy redshifted SED with the filter transmission curve, using
\begin{equation}\label{equ:convolve}
F_\lambda^{\rm mock}=\frac{\int_{-\infty}^{+\infty}S_{\lambda}\lambda R(\lambda)d\lambda}{\int_{-\infty}^{+\infty}\lambda R(\lambda)d\lambda},
\end{equation}
where $S_{\lambda}$ is the best-fit observed SED of the COSMOS/UltraVISTA galaxy,
and $R(\lambda)$ is the transmission curve of one of the 5 WFST filters.
The mock flux $F_\lambda^{\rm mock}$ is then calibrated to the mock observational flux according to the $i$ band apparent magnitude (Subaru $i^+$ flux, $F_{i^+}^{\rm obs}$) given in the COSMOS/UltraVISTA galaxy catalog.
This conversion is done by using $F_\lambda^{\rm obs}=(F_\lambda^{\rm mock}/F_{i^+}^{\rm mock})F_{i^+}^{\rm obs}$, where $F_\lambda^{\rm obs}$ is the mock observational flux, and $F_\lambda^{\rm mock}$ is the mock flux of a galaxy SED in each of the 5 WFST bands.

Dust extinction is taken into account when generating mock flux data.
The SED flux density after dust reddening from interstellar medium \citep{Calzetti_1994,Galametz_2017} can be expressed as
\begin{equation}\label{equ:extinction}
S_{\rm extinct}(\lambda_{\rm rest})=S_{\rm intrinsic}(\lambda_{\rm rest})10^{-0.4E(B-V)k(\lambda_{\rm rest})},
\end{equation}
where $E(B-V)=A_V/R_V$ is the color excess and $k(\lambda)$ is the dust extinction curve.
We adopt the \citet{Calzetti_2000} extinction curve, with
$R_V$ for this attenuation law set as 4.05.
For each galaxy, the value of attenuation $A_V$ is given by the COSMOS/UltraVISTA catalog, which is derived through the SED fitting technique.
We directly use it to generate the mock extinction-corrected fluxes.

We also consider intergalactic medium (IGM) absorption for high-redshift galaxies.
At wavelengths shorter than the ${\rm Ly}\alpha$ line, the emission can be absorbed by neutral hydrogen clouds in the IGM along our line of sight to the high-redshift galaxy.
We account for this extinction by making use of the \citet{Madau_1995} IGM attenuation law.
This is done by applying the average flux decrement $<D_A>$ between ${\rm Ly}\alpha$ and ${\rm Ly}\beta$, and $<D_B>$ between ${\rm Ly}\beta$ and the Lyman limit, such that the IGM absorption corrected flux can be written as
\begin{equation}\label{equ:absorption}
S_{\rm absorption}(\lambda_{\rm rest})=(1-<D_i>)S_{\rm initial}(\lambda_{\rm rest}) \hspace{2mm}(i=A,B),
\end{equation}
where $S_{\rm initial}$ is the initial flux density in the rest frame, adopted as the interstellar dust extinction-corrected galaxy flux $S_{\rm extinct}(\lambda_{\rm rest})$ obtained from Equation~\ref{equ:extinction}.
After these correction procedures, the galaxy SED flux density $S_{\rm absorption}$, with dust extinction and IGM absorption corrected, is substituted into Equation~\ref{equ:convolve} to generate mock flux data for all 5 WFST bands.

Next, we estimate flux errors with respect to mock WFST fluxes.
For a ground-based telescope, the signal to noise ratio (SNR) can be evaluated via the following equation \citep{Lei_2023},
\begin{equation}\label{equ:SNR}
{\rm SNR}=\frac{S\cdot A\cdot\tau}{\sqrt{S\cdot A\cdot\tau+2\cdot n_{\rm pix}\cdot[({\rm Sky}\cdot A\cdot\alpha_{\rm pix}+D)\cdot\tau+R^2]}},
\end{equation}
where $S$ is the source signal with a constant spectral flux, $\tau$ is the exposure time, $A$ is the effective area of the WFST primary mirror ($\sim4.12\times10^4~{\rm cm}^2$), $\alpha_{\rm pix}=0.111~{\rm arcsec^2}$ is the area of one pixel, $D$ is the dark current of the CCD ($D=0.005~{\rm e^-/pixel/s}$, at $-100^\circ $C), $R^2$ is the readout noise of the CCD ($R=8~{\rm e^-~rms}$), and $n_{\rm pix}$ is the total pixel number in the PSF region.
The factor of 2 applied here is because we assume that the calculation is performed on sky subtracted images.
We adopt an optimal PSF aperture of 1.18 times the full width at half maximum for a non-adaptive optics case according to the Integration Time Calculator of Gemini.
$\rm Sky$ in Equation~\ref{equ:SNR} is the sky background signal that actually lands on the detector in units of ${\rm e^-~s^{-1}~pixel^{-1}}$, which is given by
\begin{equation}\label{equ:sky}
{\rm Sky}=\int_0^\infty f_\lambda T_{\rm opt}T_{\rm band}{\rm QE_{CCD}}d\lambda,
\end{equation}
where $f_\lambda$ is the surface brightness of the sky background, $T_{opt}$ is the throughput of the optics (including the primary mirror, analog to digital converters and the 5 corrector lenses), and $\rm QE_{CCD}$ is the quantum efficiency of the CCD.

The photometric error can be evaluated through the magnitude error given by the approximate relation $\sigma_{\rm ph}\simeq2.5\log(1+1/{\rm SNR})$ \citep{Pozzetti_1998,Bolzonella_2000}.
We also add a systematic error $\sigma_{\rm sys}=0.02~{\rm mag}$ \citep{Cao_2018}, and the total magnitude error is then given by $\sigma_m=\sqrt{\sigma_{\rm ph}^2+\sigma_{\rm sys}^2}$.
Thus we can obtain the flux error $\sigma_F$ of each band from $\sigma_m$ via error propagation.
Finally, a random error drawn from the Gaussian probability distribution function (with $\sigma=\sigma_F$) is added to the mock flux in each band as the final mock photometry.

After computing and correcting for these mock fluxes, the mock observational targets obtained in the WFST shallow mode and deep mode are generated.
In this paper, we adopt 3-$\sigma$ detections to include sources into various samples as in \citet{Muzzin_2013a}, i.e.,
galaxies with fluxes that meet the $3\sigma$ depth thresholds of the 5 WFST bands (cf. Table~\ref{tab:WFST_depth}) are selected as the mock observational samples for subsequent $z_{\rm phot}$ calculation (see Section~\ref{sect:R&D}).

\section{Computation of Photometric Redshifts}\label{sect:z_compute}

\hspace{5mm}In this paper, we compute $z_{\rm phot}$ of galaxies using the mock WFST data and the ZEBRA photometric-redshift code \citep{ZEBRA} with default parameters unless stated otherwise.
The main advantage of ZEBRA is that it can generate a new set of templates adaptive to observed galaxy SEDs to minimize the mismatch between observed SEDs and available templates.
This is done by creating a training set of galaxies to optimize the shape of spectral templates that can better match predicted galaxy colors with observed ones.
We adopt the same set of 9 initial galaxy templates (see Figure~\ref{fig:templates})
as in \citet{Muzzin_2013a} for $z_{\rm phot}$ calculation using ZEBRA.
Since we have removed all the point sources that are likely bright stars or active galactic nuclei (AGNs) in the COSMOS/UltraVISTA catalog, we do not include any AGN templates during our template fitting.

First, we run ZEBRA in the photometry-check mode to identify and correct systematic errors in the photometry based on the maximum-likelihood algorithm.
ZEBRA derives a simple photometric offset in each band that minimizes the residuals between the mock observed fluxes and that of the best-fit templates, with the redshifts set as the input ones (i.e., $z_{\rm spec}$ or high-quality $z_{\rm phot}$, if $z_{\rm spec}$ are not available, from \citet{Muzzin_2013a}).
These corrections are then applied to the mock WFST photometry data, and ZEBRA iterates this procedure for 5 times to ensure that the median offset in each band converges.

Second, we run ZEBRA in the non-template-improvement mode
based on this photometric systematic offset-corrected mock catalog, using
the 9 initial galaxy templates shown in Figure~\ref{fig:templates}.
ZEBRA iteratively performs 5 logarithmic interpolations in the magnitude space
between any adjacent pair of the 9 templates, to generate $5\times8=40$ templates added to the 9 initial templates, resulting in a total of 49 templates.

Third, we run ZEBRA in the template-improvement mode, where
ZEBRA transforms the discrete template space into a linearly continuous space, using a Karhunen-Lo$\grave{e}$ve expansion to iteratively correct the eigenbases of a lower dimensional subspace through a $\chi^2$ minimization scheme.
As a result, adaptive spectral templates are generated to better match the galaxy SEDs of the training set than the set of 49 templates.

\begin{figure}
   \centering
   \includegraphics[width=0.49\columnwidth, angle=0]{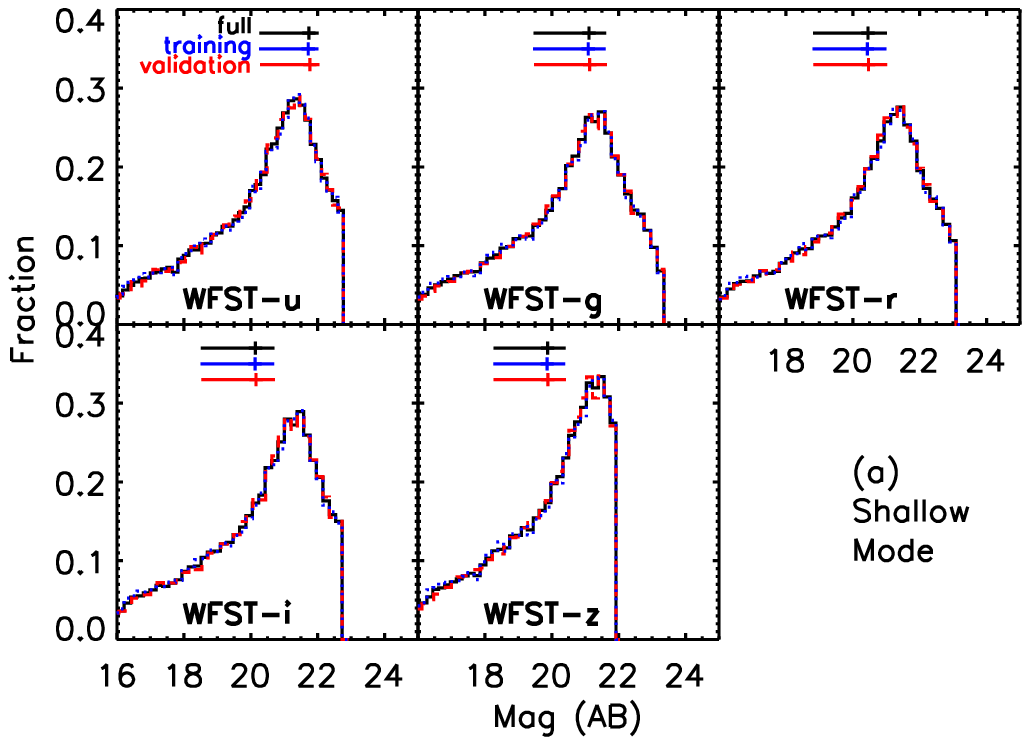}
   \includegraphics[width=0.49\columnwidth, angle=0]{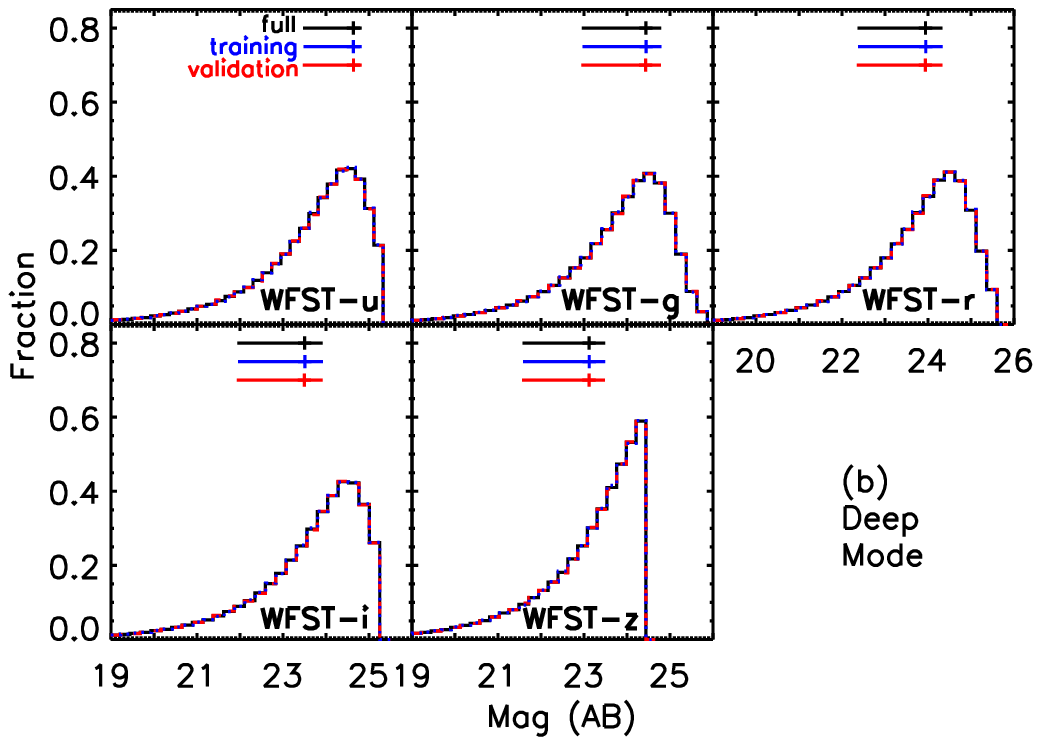}
   \caption{Magnitude distributions of the full, training, and validation mock galaxy samples (indicated by different colors) in the (a) shallow mode and (b) deep mode, respectively, with the lunar phase being fixed to 90 degree (i.e., half moon).
The color plus symbols show the medians of the distributions and the horizontal error bars indicate the 1-$\sigma$ ranges.}
\label{fig:mag_hist_compare}
\end{figure}

\begin{figure}
   \centering
   \includegraphics[width=0.49\columnwidth, angle=0]{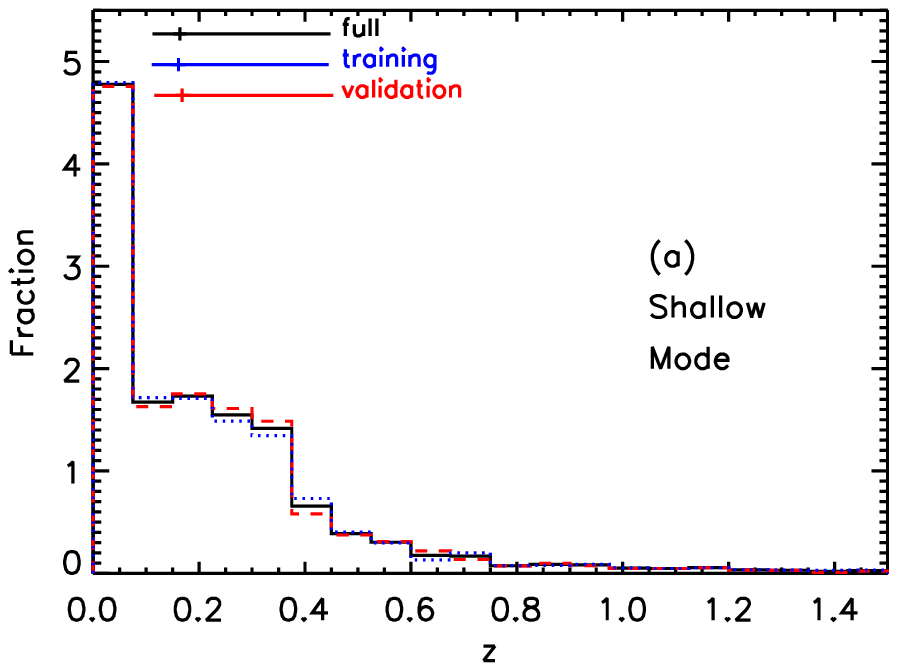}
   \includegraphics[width=0.49\columnwidth, angle=0]{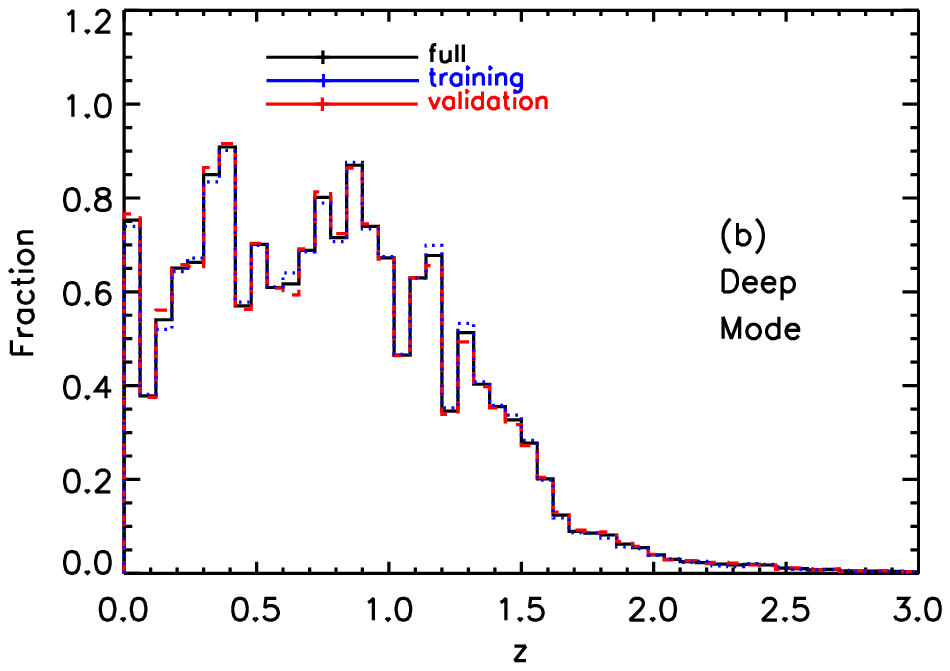}
   \caption{Distributions of input redshifts from \citet{Muzzin_2013a} of the full, training, and validation mock galaxy samples (indicated by different colors) in the (a) shallow mode and (b) deep mode, respectively, with the lunar phase being fixed to 90 degree (i.e., half moon).
The color plus symbols show the medians of the distributions and the horizontal error bars indicate the 1-$\sigma$ ranges.}
\label{fig:z_hist_compare}
\end{figure}

For each galaxy sample considered, we randomly divide all its galaxies into two equal halves: one half serves as the training set to generate a new set of adaptive templates, and $z_{\rm phot}$ computation is performed on the other half as the validation set based on these new adaptive templates plus the above 49 templates as a blind test of $z_{\rm phot}$ quality.
We compare multiwavelength photometry and input redshifts of both the training and validation sets in Figures~\ref{fig:mag_hist_compare} and \ref{fig:z_hist_compare}, respectively.
We find that
they have almost identical photometric and redshift properties, such that the templates generated based on the randomly selected training set of galaxies can be adaptive to the full galaxy sample, and that $z_{\rm phot}$ computation on the validation set as a blind test can be representative of the result for the full sample.

In the template-improvement mode, ZEBRA iterates twice, i.e., over the redshift of 0--3 as one single bin and in smaller redshift bins of 0.5, to train the 49 templates based on a chosen training set.
Narrowing down the redshift bin (e.g., to $\Delta z=0.2$) only increases the total number of adaptive templates generated, but has little effect on the final $z_{\rm phot}$ results.
Therefore, we use a total of $49\times6+49=343$ final templates and run ZEBRA to compute $z_{\rm phot}$ for each selected galaxy sample.

\section{Results and Discussion}\label{sect:R&D}

\hspace{5mm}In this section, we show $z_{\rm phot}$ results with mock WFST data
in the shallow and deep modes given various lunar phases (see Sections~\ref{sect:R&D_shallow} and \ref{sect:R&D_deep}, respectively),
compare our WFST $z_{\rm phot}$ results with that from some recent works (see Section~\ref{sect:R&D_com}),
and assess the improvement of $z_{\rm phot}$ quality with the addition of other data (see Section~\ref{sect:R&D_CSST_Euclid}).

\subsection{$z_{\rm phot}$ results in the shallow mode}\label{sect:R&D_shallow}

\begin{figure}
   \centering
   \vspace{-2cm}
   \includegraphics[width=0.98\columnwidth, angle=0]{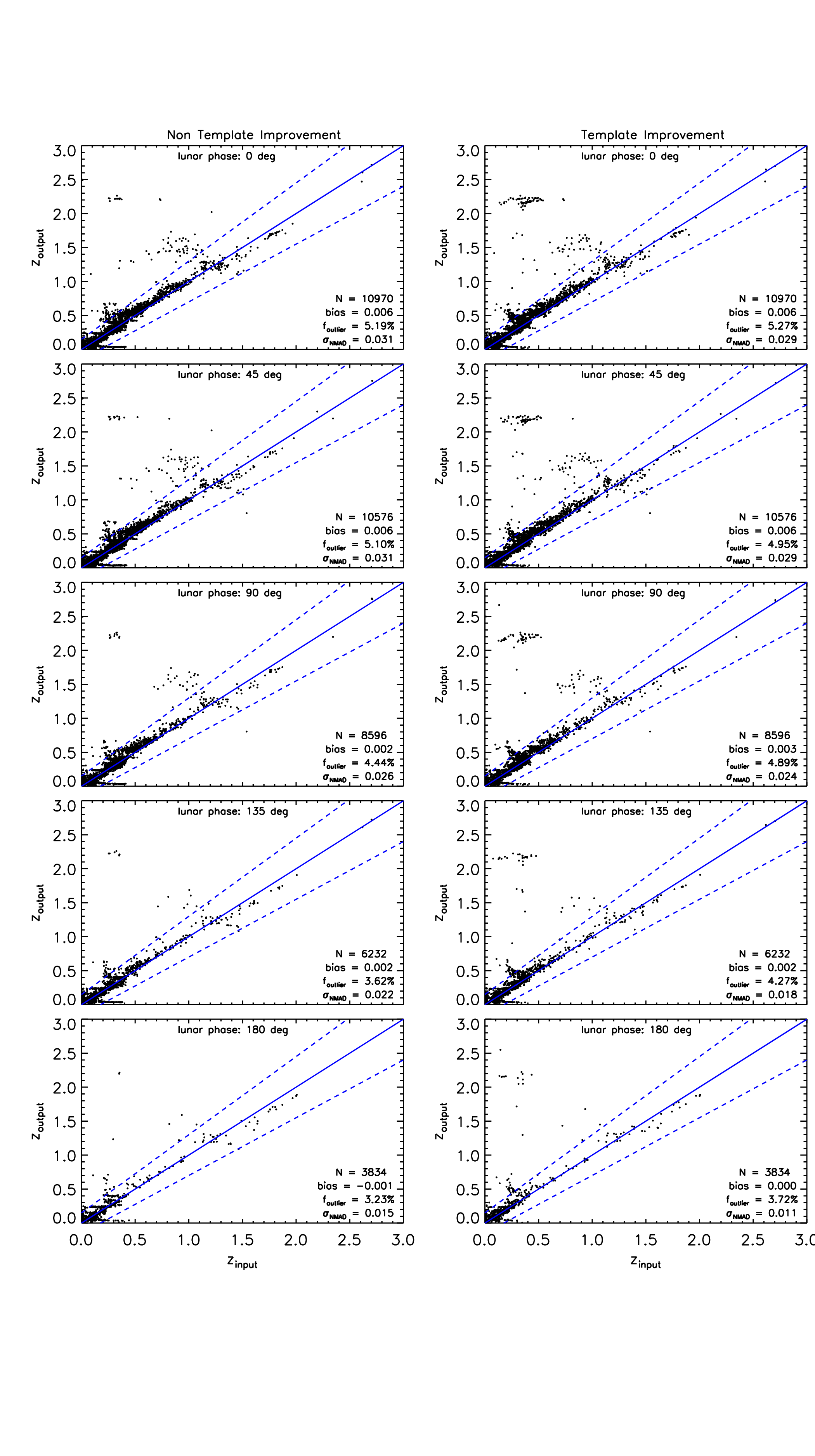}
   \vspace{-3cm}
   \caption{$z_{\rm phot}$ results in the shallow mode
with ZEBRA run in the non-template-improvement mode (left panels) and template-improvement mode (right panels) under various lunar phases ($\rm 0~deg$: no moon; $\rm 45~deg$: 1/4 moon; $\rm 90~deg$: half moon; $\rm 135~deg$: 3/4 moon; and $\rm 180~deg$: full moon), respectively.
In each panel, blue dashed lines depict the boundary of $z_{\rm phot}$ outliers, and the
number of sources considered, bias, $f_{\rm outlier}$, and $\sigma_{\rm NMAD}$ are annotated.}
\label{fig:zcom_shallow}
\end{figure}

\hspace{5mm}The $z_{\rm phot}$ results with mock WFST data
in the shallow mode are shown in Figure~\ref{fig:zcom_shallow},
whose left and right panels are for the non-template-improvement and template-improvement modes under various lunar phases, respectively.
To evaluate $z_{\rm phot}$ quality, we adopt some commonly-used quantities:
(1) normalized median absolute deviation \citep[e.g.,][]{EAZY}, i.e., $\sigma_{\rm NMAD}=1.48\times(\left|\frac{\Delta z-{\rm median}(\Delta z)}{1+z_{\rm input}}\right|)$,
where $\Delta z=z_{\rm output}-z_{\rm input}$,
with $z_{\rm output}$ and $z_{\rm input}$ being the output $z_{\rm phot}$ and
input redshifts from the COSMOS/UltraVISTA catalog \citep{Muzzin_2013a}, respectively;
(2) outlier fraction $f_{\rm outlier}$, with outliers being defined as sources with $\left|\Delta z\right|/(1+z_{\rm input})>0.15$; and
(3) bias, i.e., median of $\Delta z/(1+z_{\rm input})$ with outliers being removed.

According to Figure~\ref{fig:zcom_shallow}, under various lunar phases,
we have bias=$-0.001$--0.006, $\sigma_{\rm NMAD}=0.015$--0.031, and $f_{\rm outlier}=3.23\%$--5.19\% in the non-template-improvement mode, and have
bias=$0.000$--0.006, $\sigma_{\rm NMAD}=0.011$--0.029, and $f_{\rm outlier}=3.72\%$--5.27\% in the template-improvement mode, respectively.
The template-improvement mode delivers smaller biases and $\sigma_{\rm NMAD}$ than
the non-template-improvement mode, which is expected;
however, the former mode provides comparable or even slightly larger $f_{\rm outlier}$ than
the latter mode, due to misidentification of Lyman break as Balmer break or vice versa
that is caused by the relatively limited photometry (i.e., only $ugriz$ bands)
although the significantly enlarged template set can cover the full parameter space of the observed galaxy SEDs.

\begin{figure}
   \centering
   \includegraphics[width=0.49\columnwidth, angle=0]{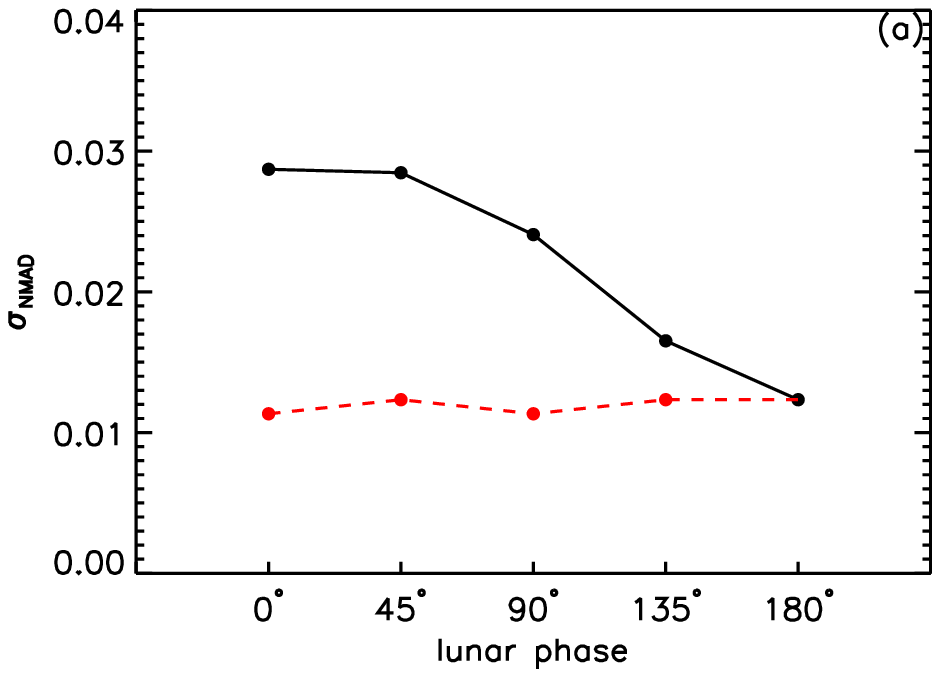}
   \includegraphics[width=0.49\columnwidth, angle=0]{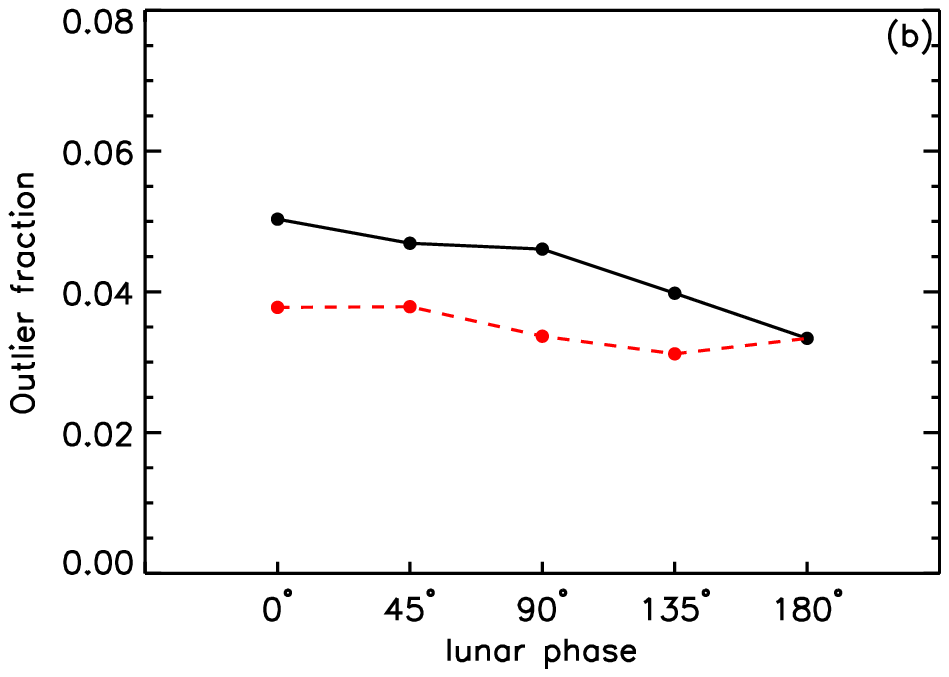}
   \includegraphics[width=0.49\columnwidth, angle=0]{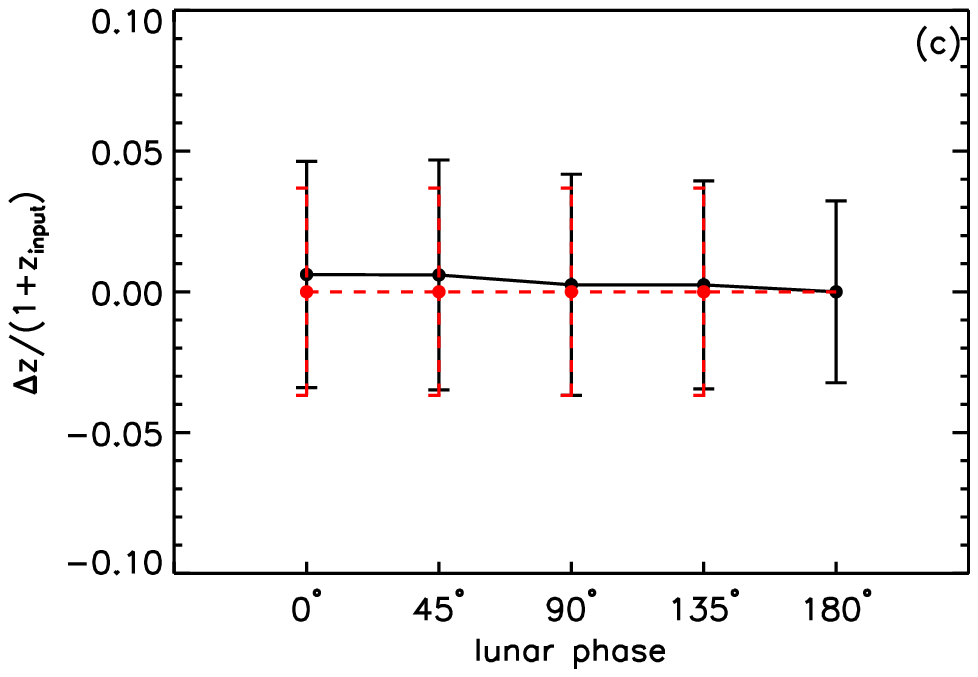}
   \includegraphics[width=0.49\columnwidth, angle=0]{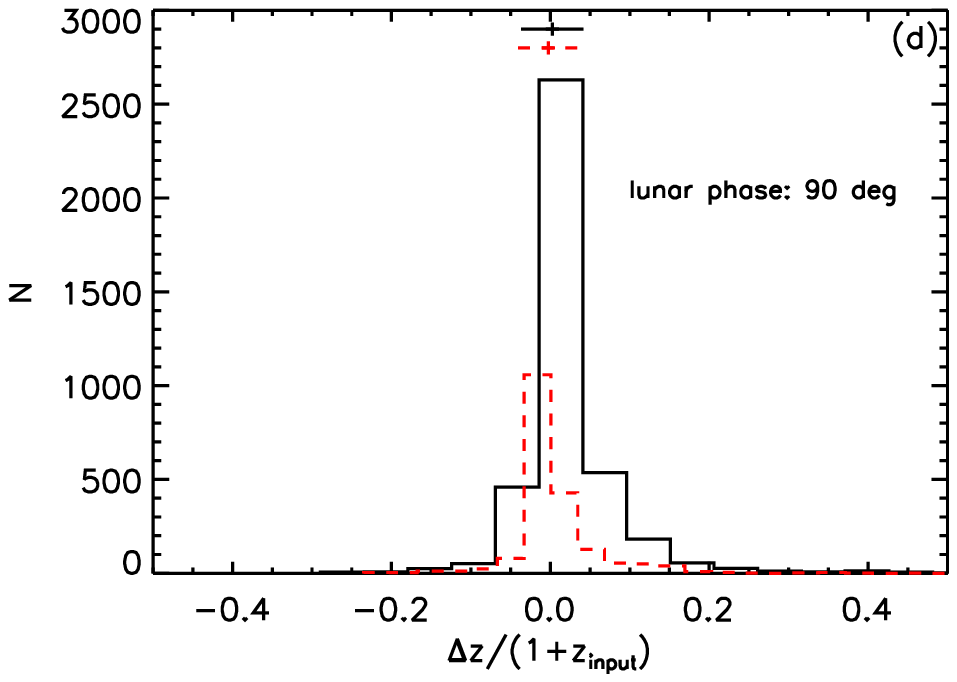}
   \caption{Dependences of $\sigma_{\rm NMAD}$, $f_{\rm outlier}$, and bias on lunar phase as well as distribution of $\Delta z/(1+z_{\rm input})$ in the shallow mode, with the full mock sample and the specific sample of galaxies observable under full moon indicated by the black and red dashed lines and histogram, respectively.
In the bottom-right panel, the lunar phase is fixed to $\rm 90~deg$; the plus sign and its horizontal error bar show the median and 1-$\sigma$ range of $\Delta z/(1+z_{\rm input})$.}
\label{fig:moonphase_shallow}
\end{figure}

As shown in Figure~\ref{fig:moonphase_shallow}, $z_{\rm phot}$ quality shows some variation with lunar phase: $z_{\rm phot}$ quality improves as the lunar phase increases, with the best $z_{\rm phot}$ result achieved under the lunar phase of $\rm 180~deg$ (full moon).
Two factors can influence $z_{\rm phot}$ quality of the selected sample under different lunar phases.
One is the lunar phase itself: under brighter lunar phases,
the sky light background contributed by the moon becomes larger, resulting in
larger uncertainties on photometry and eventually worse $z_{\rm phot}$ quality.
The other is sample selection effect:
under brighter lunar phases, only brighter sources can be well observed,
which usually have higher-SNR photometry that leads to higher-quality $z_{\rm phot}$.

To make a more sensible evaluation of the lunar phase influence and consider the above two factors separately, we restrict the sample observable under full moon and measure $z_{\rm phot}$ under different lunar phases, with the results shown as the red dashed lines in Figure~\ref{fig:moonphase_shallow}.
It is clear that lunar phase has a very limited influence on $z_{\rm phot}$ results of a fixed sample.
Therefore, the variation in $z_{\rm phot}$ quality under different lunar phases in the shallow mode is primarily driven by sample-selection effect.

\subsection{$z_{\rm phot}$ results in the deep mode}\label{sect:R&D_deep}

\begin{figure}
   \centering
   \includegraphics[width=0.49\columnwidth, angle=0]{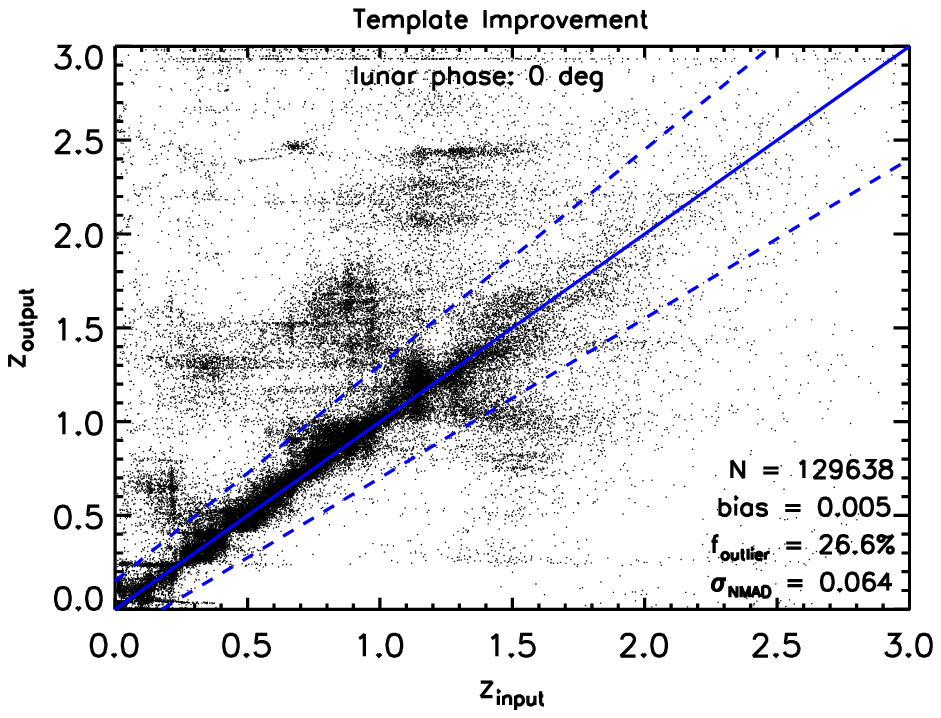}
   \includegraphics[width=0.49\columnwidth, angle=0]{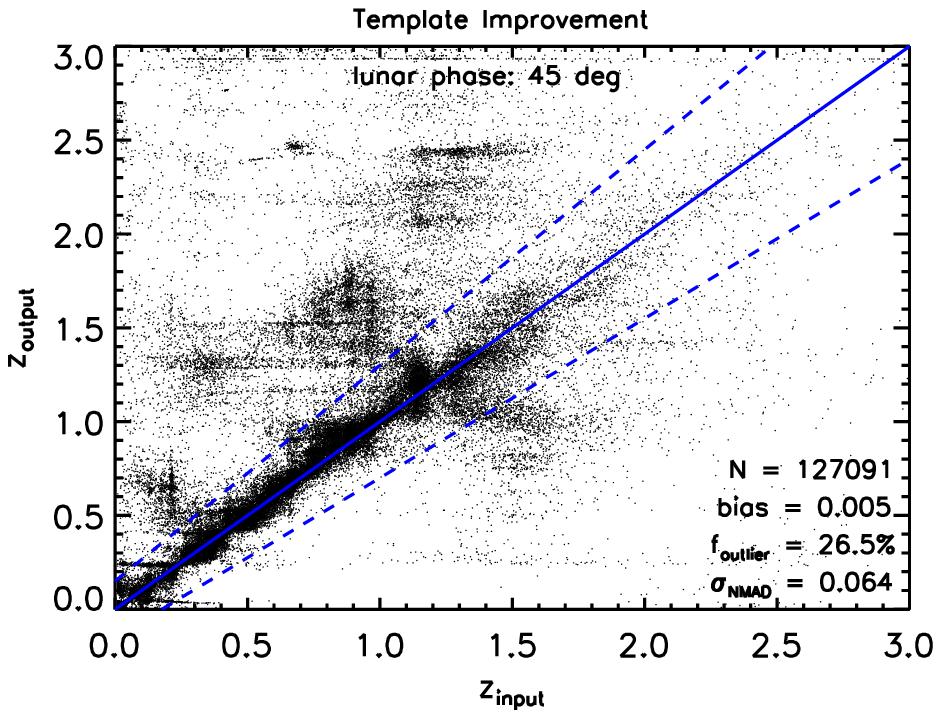}
   \includegraphics[width=0.49\columnwidth, angle=0]{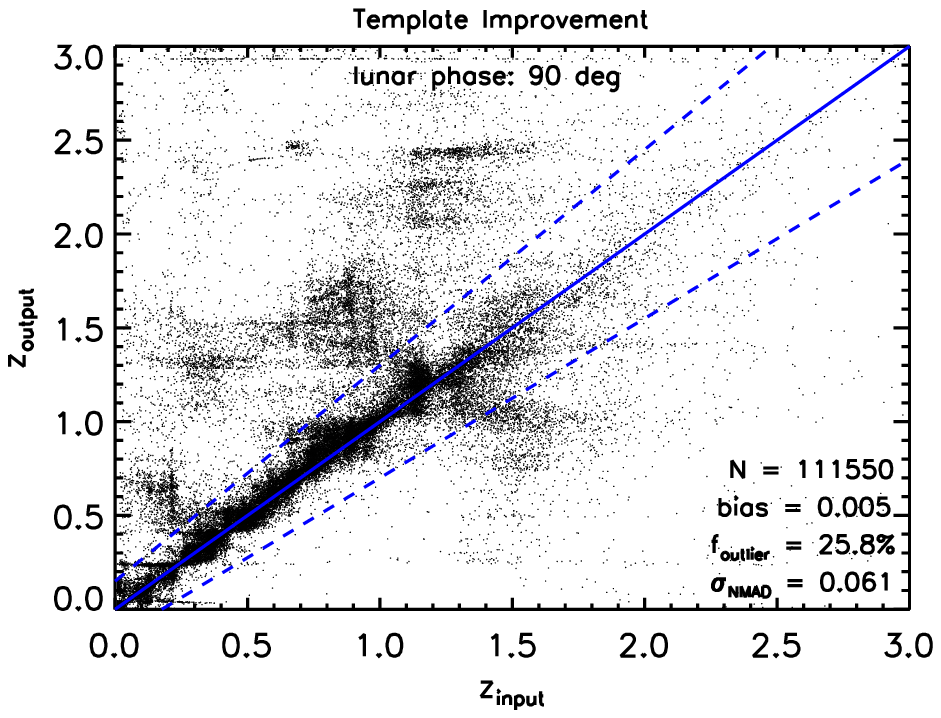}
   \includegraphics[width=0.49\columnwidth, angle=0]{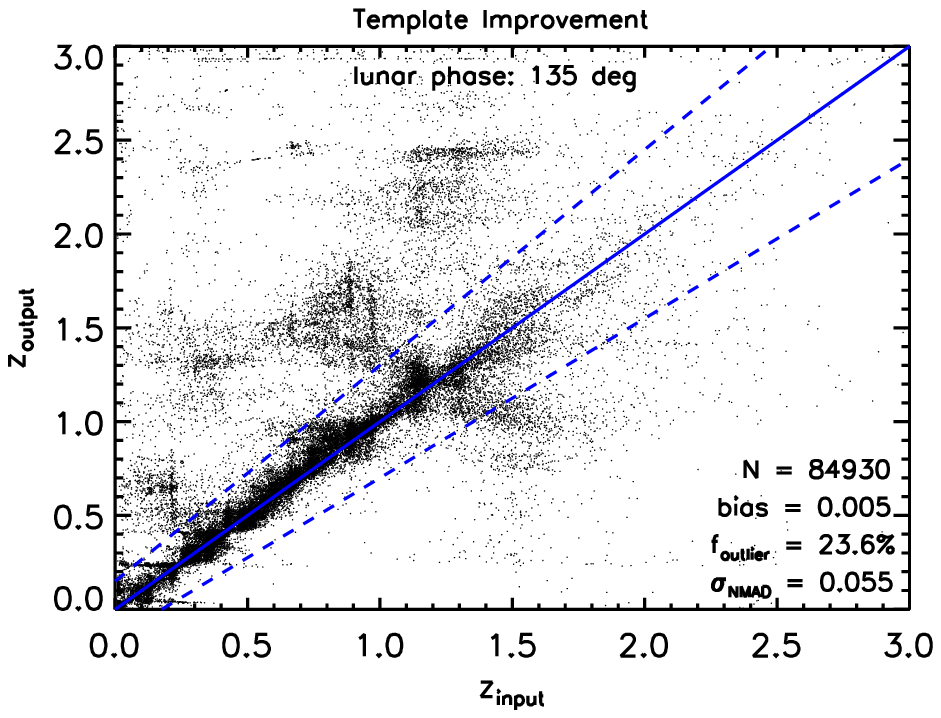}
   \includegraphics[width=0.49\columnwidth, angle=0]{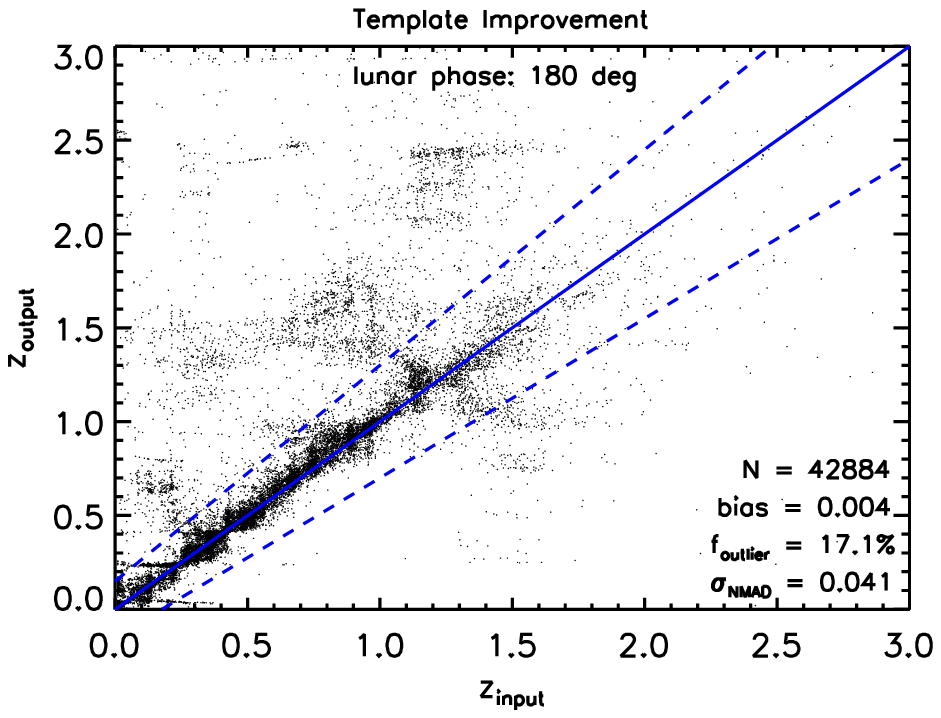}
   \caption{Similar to Figure~\ref{fig:zcom_shallow}, but for
$z_{\rm phot}$ results in the deep mode
with ZEBRA run in the template-improvement mode under various lunar phases.}
\label{fig:zcom_deep}
\end{figure}

\hspace{5mm}The $z_{\rm phot}$ results with mock WFST data in the deep mode
under various lunar phases,
with ZEBRA run in the template-improvement mode,
are show in Figure~\ref{fig:zcom_deep}.
Apparently, the inclusion of large amounts of faint galaxies significantly reduces $z_{\rm phot}$ quality: $\sigma_{\rm NMAD}$ grows from 0.041 to 0.064 with the dimming of lunar phase;
$f_{\rm outlier}$ increases to $26.6\%$ when there is no moon;
bias$\sim0.005$, being almost constant and comparable to the situations of faint lunar phases in the shallow mode.

\begin{figure}
   \centering
   \includegraphics[width=0.49\columnwidth, angle=0]{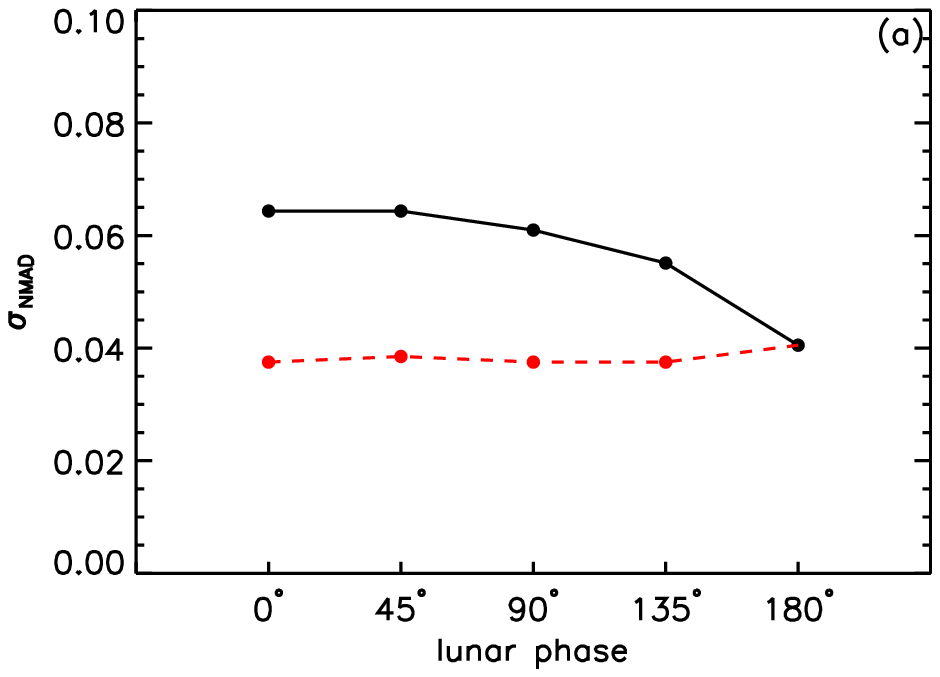}
   \includegraphics[width=0.49\columnwidth, angle=0]{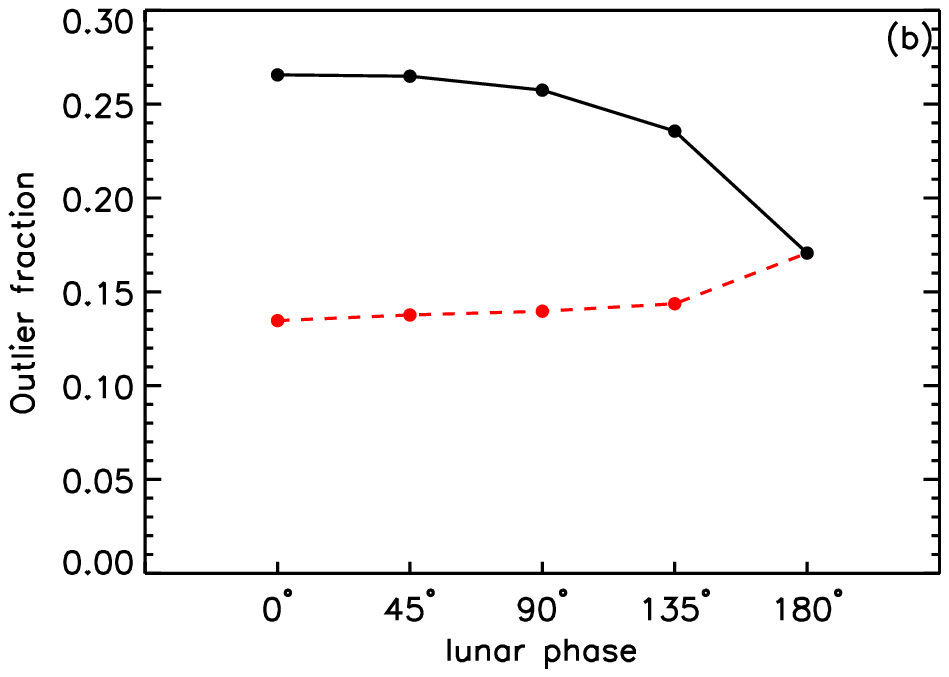}
   \includegraphics[width=0.49\columnwidth, angle=0]{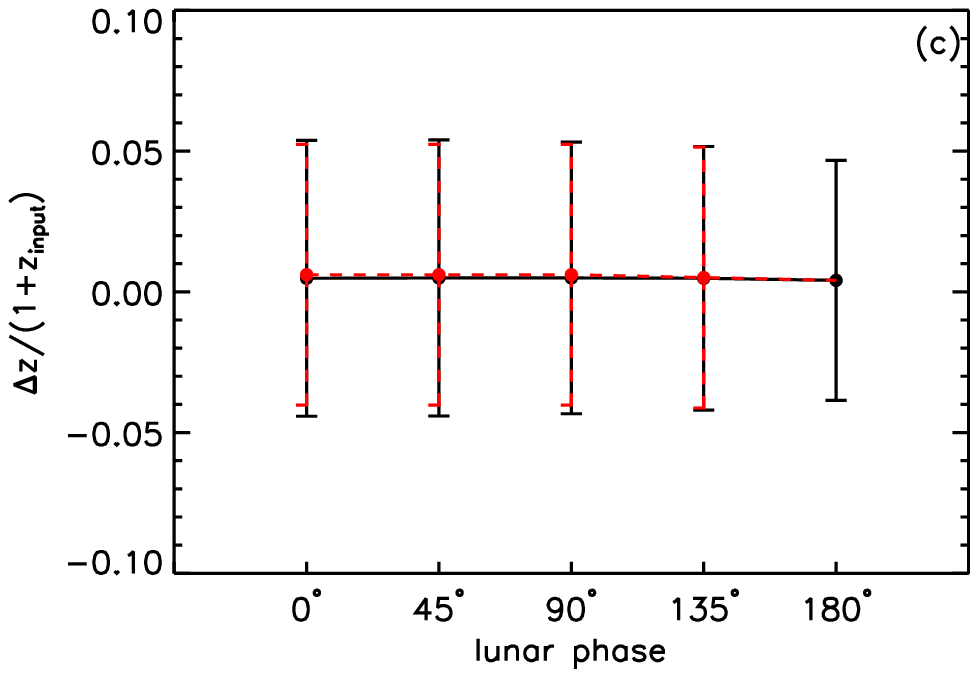}
   \includegraphics[width=0.49\columnwidth, angle=0]{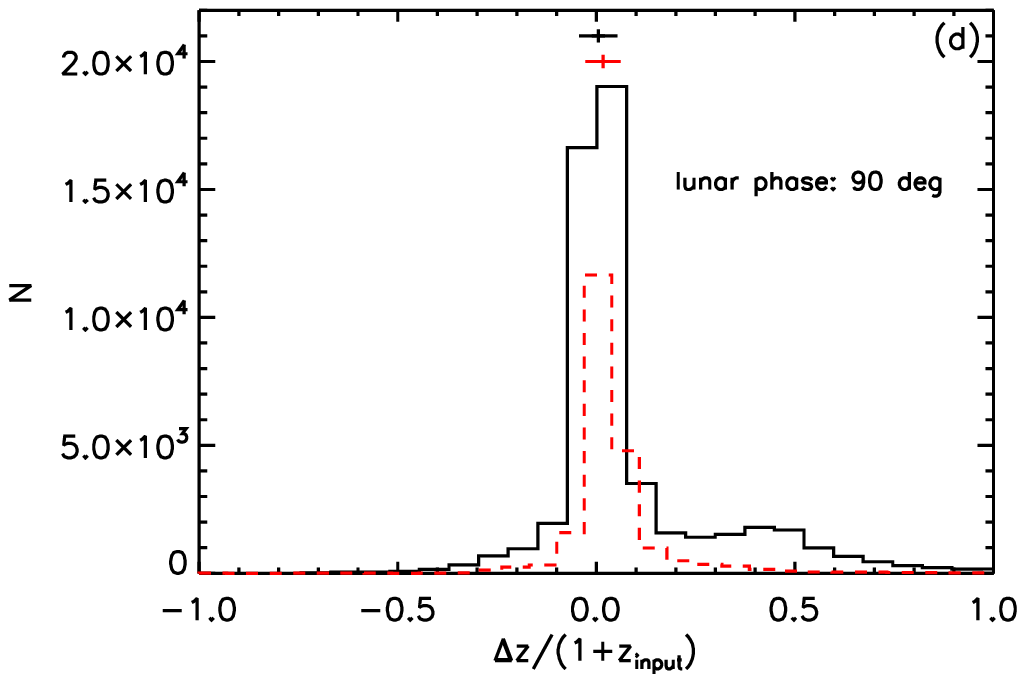}
   \caption{Same as Figure~\ref{fig:moonphase_shallow},
but for dependences of $\sigma_{\rm NMAD}$, $f_{\rm outlier}$, and bias on lunar phase as well as distribution of $\Delta z/(1+z_{\rm input})$ in the deep mode.}
\label{fig:moonphase_deep}
\end{figure}

In the deep mode, $z_{\rm phot}$ quality shows a stronger variation with lunar phase than in the shallow mode, as shown in Figure~\ref{fig:moonphase_deep}.
However, this does not mean that dimming of moonlight will cause $z_{\rm phot}$ quality to decrease for a fixed galaxy sample.
When we consider the fixed sample of galaxies observable under full moon,
we find that dimming of sky background caused by moonlight slightly reduces photometric uncertainties and thus improves $z_{\rm phot}$ quality, e.g., $f_{\rm outlier}$ decreasing from $17.1\%$ (full moon) to $\leq14\%$ (no moon) (see the red dashed lines in Figure~\ref{fig:moonphase_deep}).
Thus, the downgrade of $z_{\rm phot}$ quality under fainter lunar phases in the deep mode is a direct result of sample-selection effect, same as in the shallow mode. The dimmer moonlight in the deep mode enables the detection of fainter populations of galaxies, which often exhibit poorer photometry qualities; consequently, this leads to a continuous decrease in the accuracy and reliability of $z_{\rm phot}$ estimation.
Therefore, we conclude that lunar phase only has negligible or very slight effects on $z_{\rm phot}$ quality for a given sample of galaxies; however, it can have a strong influence on sample selection, resulting in ``apparent'' variation of $z_{\rm phot}$ quality across different samples.

\subsection{Comparison with other $z_{\rm phot}$ results}\label{sect:R&D_com}

\begin{figure}
   \centering
   \includegraphics[width=0.49\columnwidth, angle=0]{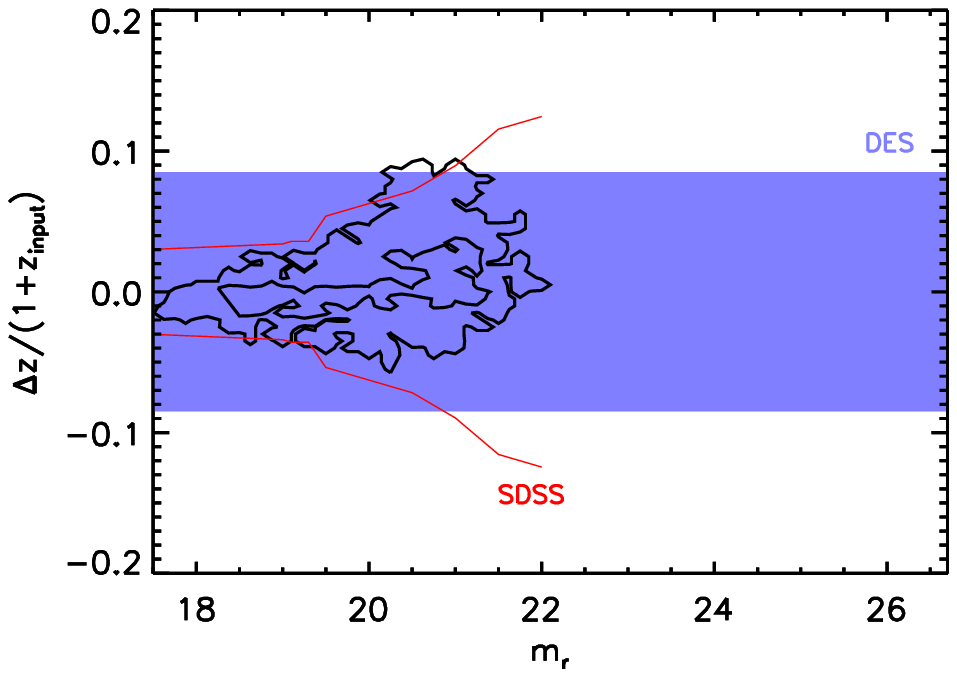}
   \includegraphics[width=0.49\columnwidth, angle=0]{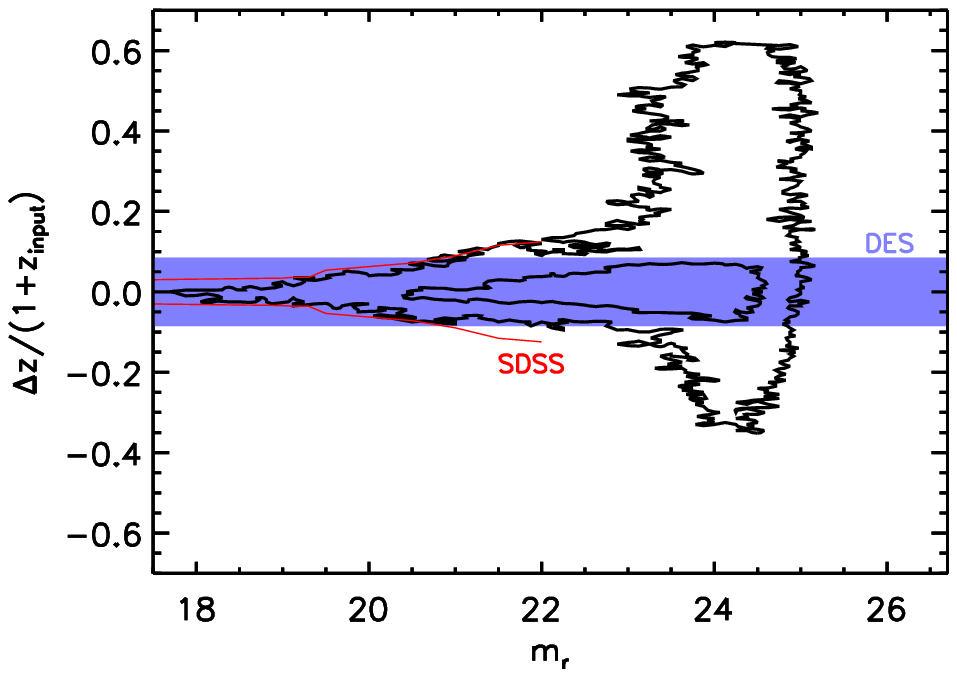}
   \caption{$\Delta z/(1+z_{\rm input})$ as a function of $r$-band magnitude
in the shallow mode (left) and deep mode (right), respectively.
The black envelopes show the 2-$\sigma$ and 3-$\sigma$ contours surrounding the peak distributions.
For comparison, the red curves show the derived average cumulative rms deviation of SDSS galaxies based on $ugriz$-band photometry as a function of $r$-band magnitude \citep{Csabai_2003};
the blue horizontal bars indicate the single-value (i.e., derived with the entire sample) standard deviation of DES galaxies based on $grizY$-band photometry \citep{Yang_2023}.}
\label{fig:dz_R}
\end{figure}

\hspace{5mm}We compare our WFST $z_{\rm phot}$ results with some relevant works;
for simplicity, we fix the lunar phase in the WFST mock data to $\rm 90~deg$ (half moon) here.
Figure~\ref{fig:dz_R} shows $\Delta z/(1+z_{\rm input})$ as a function of $r$-band magnitude in the shallow mode and deep mode (see the black contours), respectively.
Overall, the bright sources in the shallow mode have much better $z_{\rm phot}$ than the faint sources in the deep mode, with the scatter of $\Delta z/(1+z_{\rm input})$ of the latter being $\sim2$--3 times larger than that of the former.
The red curves in Figure~\ref{fig:dz_R} show the average cumulative rms deviation between $z_{\rm phot}$ and $z_{\rm spec}$ as a function of $r$-band magnitude in the SDSS survey early data release \citep[using the $ugriz$-band photometry;][]{Csabai_2003}, where $z_{\rm phot}$
were derived with a hybrid technique (empirical and template fitting methods) to calibrate galaxy SED templates to improve $z_{\rm phot}$ quality,
utilizing a training set of galaxies with secure $z_{\rm spec}$.
We find that, at $m_r<22$, our $\Delta z/(1+z_{\rm input})$ scatter is generally comparable to or smaller than that of \citet{Csabai_2003}.
This is partly because their training set of galaxies are restricted to the bright population, which makes it difficult to constrain $z_{\rm phot}$ scatter toward the faint end.
Recently, \citet{Yang_2023} estimated $z_{\rm phot}$ of galaxies and quasars in the Southern Hemisphere DES wide survey based on a Bayesian analysis algorithm in the multi-color space, using the $grizY$-band photometry.
We show the standard deviation of $\Delta z/(1+z_{\rm input})$ of their galaxies in the blue bars in Figure~\ref{fig:dz_R}, which is comparable to our result in the shallow mode.

\begin{figure}
   \centering
   \includegraphics[width=0.49\columnwidth, angle=0]{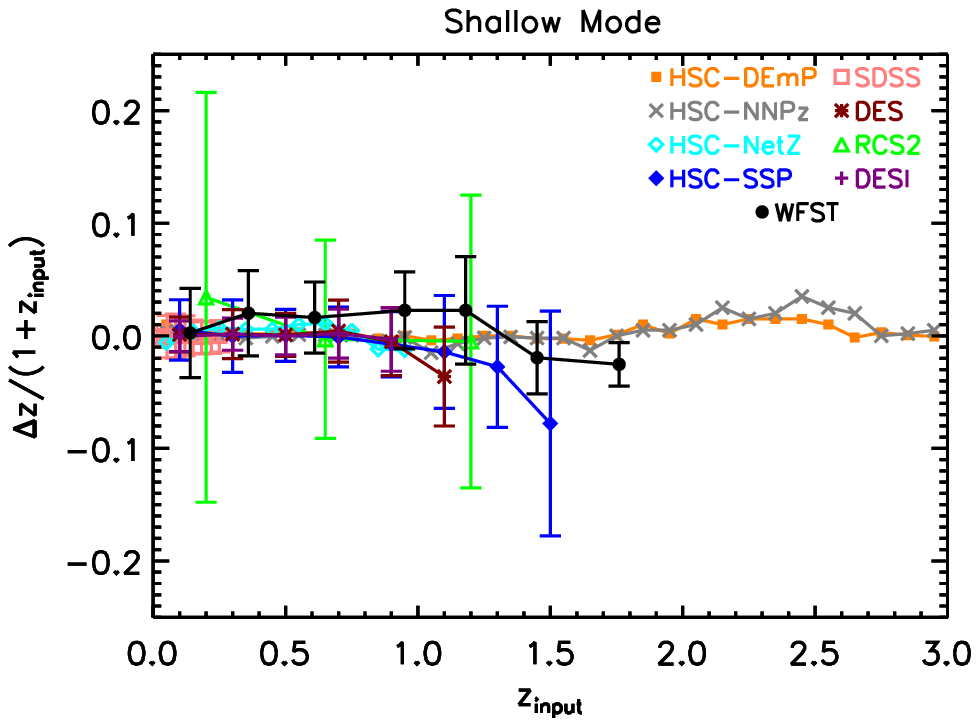}
   \includegraphics[width=0.49\columnwidth, angle=0]{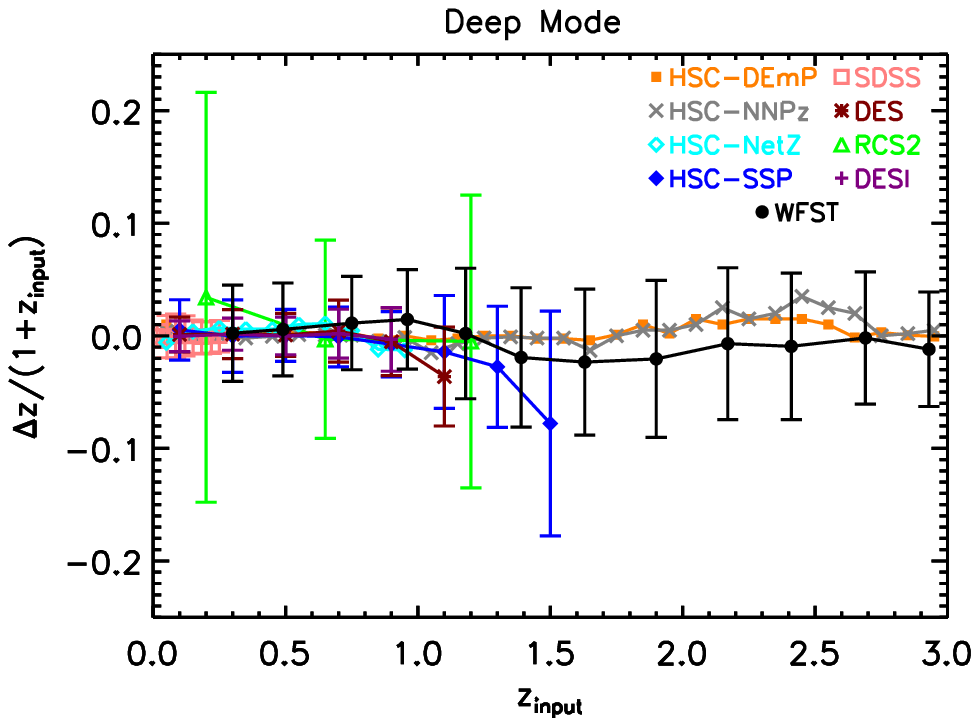}
   \caption{$\Delta z/(1+z_{\rm input})$ as a function of $z_{\rm input}$ in the shallow mode (left) and deep mode (right), respectively.
Also shown for comparison are those from the HSC survey (using convolutional neural network for $z_{\rm phot}$ computation, NetZ; cyan; \citealt{Schuldt_2021}), second Red-Sequence Cluster Survey (using Direct Empirical Photometric method, DEmP; green; \citealt{Hsieh_2014}), HSC-SSP survey (using K Nearest Neighbor, KNN; blue; \citealt{Zou_2022}), DES survey (using KNN; brown; \citealt{Zou_2022}), DESI survey (using KNN; purple; \citealt{Zou_2022}), HSC survey (using DEmP; orange; \citealt{Tanaka_2018}; no error bars provided), HSC survey (using Nearest Neighbor, NNPz; gray; \citealt{Tanaka_2018}; no error bars provided), and SDSS survey (using random forest regression; pink; \citealt{Carliles_2010}).}
\label{fig:dz_z}
\end{figure}

Figure~\ref{fig:dz_z} shows $\Delta z/(1+z_{\rm input})$ as a function of $z_{\rm input}$ in the shallow mode and deep mode, respectively, in comparison with several other works.
In general, our $\Delta z/(1+z_{\rm input})$ shows a smooth distribution in each smaller redshift bin; the biases and scatters of our $z_{\rm phot}$ are smaller than many quoted results from other works up to $z\sim3$.
This may be because the training sets we use to improve the galaxy SED templates are randomly selected, thereby having good coverage of various galaxy properties and being representative of the full galaxy sample (see Figures~\ref{fig:mag_hist_compare} and \ref{fig:z_hist_compare}).
However, in real observations, $z_{\rm spec}$ of the training sets would be mostly limited to bright sources and low redshifts, being difficult to well cover the full properties of the selected galaxy sample; in addition, the observed galaxy SEDs can be very different from the galaxy templates adopted here; therefore,
a nonnegligible effect on actual biases and scatters of our $z_{\rm phot}$ in real observations would be expected.

\begin{figure}
   \centering
   \includegraphics[width=0.49\columnwidth, angle=0]{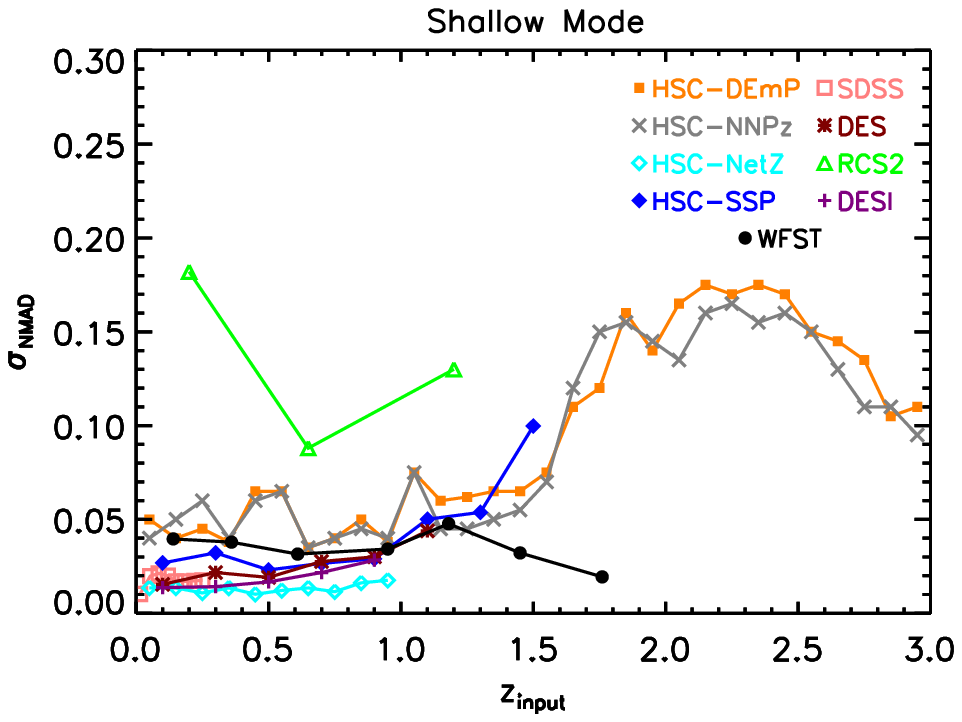}
   \includegraphics[width=0.49\columnwidth, angle=0]{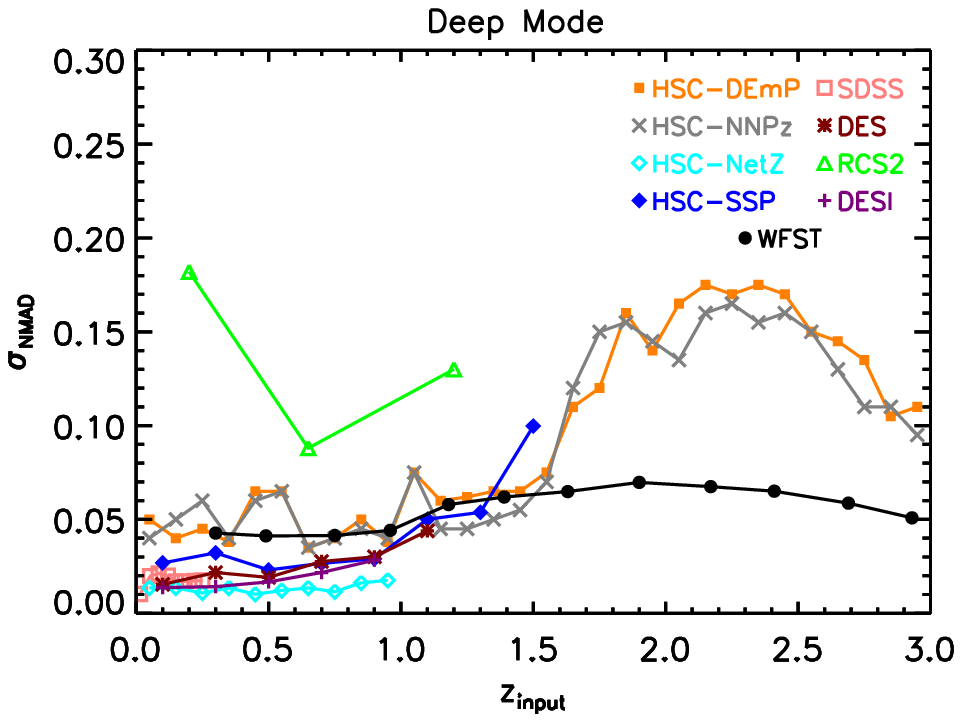}
   \includegraphics[width=0.49\columnwidth, angle=0]{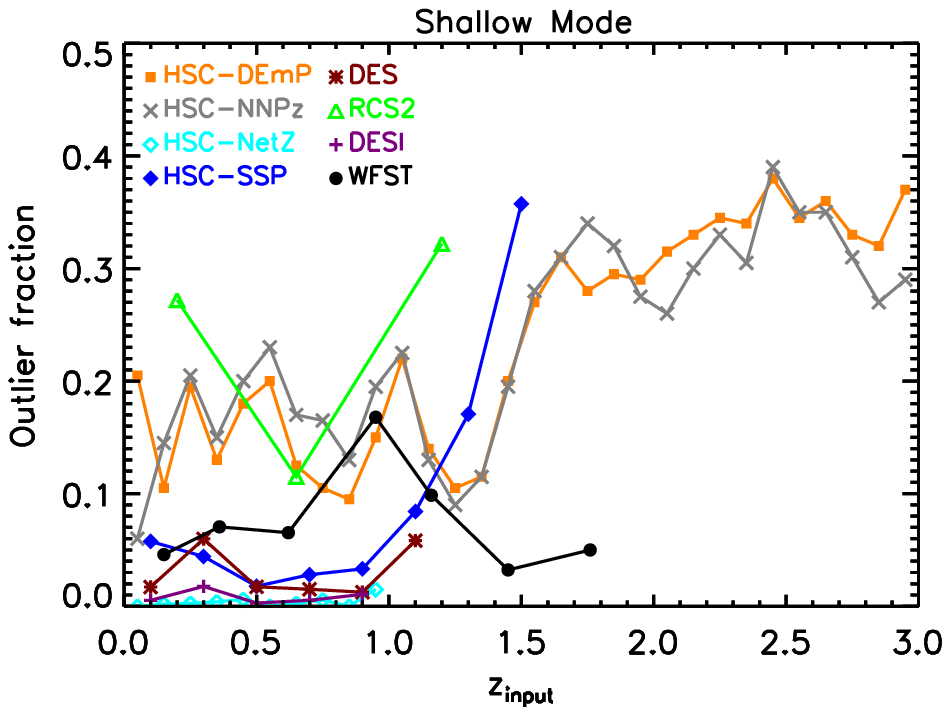}
   \includegraphics[width=0.49\columnwidth, angle=0]{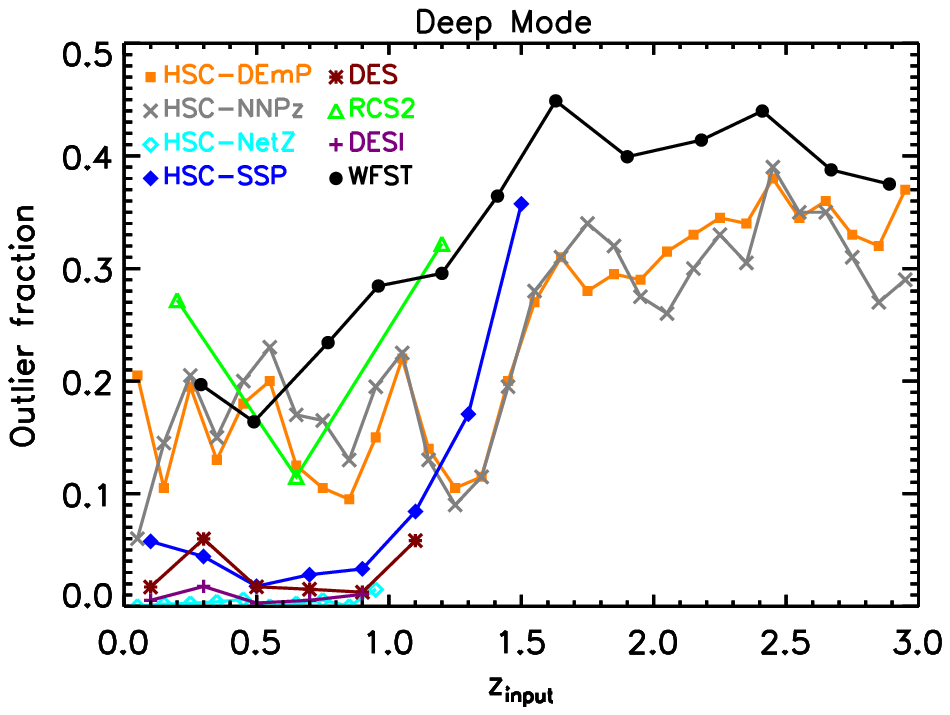}
   \caption{$\sigma_{\rm NMAD}$ and $f_{\rm outlier}$ as a function of $z_{\rm input}$ in the shallow mode (left) and deep mode (right), respectively.
The comparison surveys are the same as those in Figure~\ref{fig:dz_z}.}
\label{fig:sigma_outlier_z}
\end{figure}

Figure~\ref{fig:sigma_outlier_z} shows
$\sigma_{\rm NMAD}$ and $f_{\rm outlier}$ as a function of $z_{\rm input}$ in the shallow mode and deep mode, respectively, in comparison with the aforementioned works.
Again, our $z_{\rm phot}$ results are overall in line with those in the literature.
At $z<1.5$, our $z_{\rm phot}$ quality is comparable to those of most other works,
but not better than those works based on machine deep learning, e.g., using random forest algorithms or convolutional neural networks.
At $z\ge1.5$, both our $z_{\rm phot}$ results and the quoted results deteriorate;
our $f_{\rm outlier}$ is larger than the results based on the 5-band HSC photometry
that includes the near-infrared $Y$ band conducive to $z_{\rm phot}$ improvement at high redshifts.
In contrast, our $\sigma_{\rm NMAD}$ remains largely constant and acceptably small both in the shallow mode and deep mode, and within the full redshift range of 0--3 explored here.

It is clear that at low redshifts ($z<1.5$), to a certain degree, the machine deep learning procedures can effectively further improve $z_{\rm phot}$ results compared to the traditional template-fitting techniques, which is usually done by applying a large training sample with secure $z_{\rm spec}$ and high-quality observed SEDs.
At higher redshifts ($z\ge1.5$), however, such a training sample would become very incomplete, which makes it difficult to cover the full parameter space of all observed sources;
thus, it is still unlikely to precisely constrain uncertainties of $z_{\rm phot}$ measurement at high redshifts simply based on machine learning.
At $z\ge1.5$, the traditional template-fitting technique still shows advantages in some respects, e.g., as shown in Figure~\ref{fig:sigma_outlier_z}, our $\sigma_{\rm NMAD}$ outperform that of machine-learning results,
because ZEBRA can extend the known templates in the multi-parameter space and improve the fitting result by creating new templates and optimizing their shapes to be adaptive to galaxy multiwavelength photometry.
However, the ZEBRA template-improvement procedure does not seem to effectively reduce $f_{\rm outlier}$ at $z\ge1.5$, mainly due to misidentification of spectral breaks or other spectral features in galaxy SEDs thanks to the limited $ugriz$-band photometry.
In contrast, the most recent machine learning methods based on the Direct Empirical Photometric (DEmP) or Nearest Neighbor (NNPz) method seem to have the potential to reduce $f_{\rm outlier}$ to a large extent.
Therefore, in the future, we can combine the machine learning methods with adaptive template fitting procedures to further improve WFST $z_{\rm phot}$ quality.

\subsection{Improvement of $z_{\rm phot}$ quality with the addition of other data}\label{sect:R&D_CSST_Euclid}

\hspace{5mm}We further investigate the improvement of WFST $z_{\rm phot}$ quality by including mock data from the China Space Station Telescope \citep[{\it CSST, to be launched around 2024};][]{Zhan_2011} and {\it Euclid} space telescope \citep[launched in July 2023;][]{Laureijs_2012}, both of which can provide additional high-quality ultraviolet and/or near-infrared data in large sky areas that are critically supplementary to WFST data.

We consider the {\it CSST} $NUV$- and $y$-band mock data,
whose photometric errors are measured via SNR \citep{Ubeda_2011}:
\begin{equation}\label{equ:SNR_CSST}
{\rm SNR} = \frac{C_st}{\sqrt{C_st+N_{\rm pix}(B_{\rm sky}+B_{\rm det})t+N_{\rm pix}N_{\rm read}R_n^2}},
\end{equation}
where $t$ is the exposure time and $N_{\rm pix}$ is the number of detector pixels covered by a source.
$N_{\rm pix}$ is 16 by default, corresponding to the case of a point source in the image;
changing $N_{\rm pix}$ value does not significantly alter the final result.
$N_{\rm read}$ is the number of detector readouts, $B_{\rm det}$ is the detector dark current, and $R_{\rm n}$ is the
read noise.
Default parameter settings of $t= 300~$s, $N_{\rm read}=2$, $B_{\rm det}=0.02~e^-~s^{-1}~{\rm pixel}^{-1}$, and $R_{\rm n}=5~e^-~{\rm pixel}^{-1}$ are adopted.
$C_s$ is the count rate from the source in units of $e^-~s^{-1}$.
$B_{\rm sky}$ in Equation~\ref{equ:SNR_CSST} is the sky background in $e^-~s^{-1}~{\rm pixel}^{-1}$.
For more details about the CSST mock flux and error estimation, we refer readers to Section 2.3 in \citet{Cao_2018}.

We consider the {\it Euclid} $Y_{\rm E}$-, $J_{\rm E}$- and $H_{\rm E}$-band mock data.
Since we do not have specific details of {\it Euclid} (such as those of {\it CSST} shown in Equation~\ref{equ:SNR_CSST}), we adopt photometric errors in the similar $Y$, $J$, and $H$ bands of the VISTA survey for approximation, i.e.,
photometric errors of mock {\it Euclid} data are directly taken from the \citet{Muzzin_2013a} catalog, which are scaled proportionally to mock $Y_{\rm E}$-, $J_{\rm E}$- and $H_{\rm E}$-band fluxes.
Given that there is a slight bias between the ground-based VISTA telescope and {\it Euclid}, we apply a constant conversion factor to convert the VISTA errors to the {\it Euclid} mock errors, which is defined as the ratio of flux error between the {\it CSST} $y$ band and VISTA $Y$ band for each source at the given magnitude.
We then compute mock fluxes and flux errors in the {\it CSST} $NUV$, $y$ and {\it Euclid} $Y_{\rm E}$, $J_{\rm E}$ and $H_{\rm E}$ bands, which are subsequently combined with WFST mock data for $z_{\rm phot}$ improvement.

\begin{figure}
   \centering
   \includegraphics[width=0.49\columnwidth, angle=0]{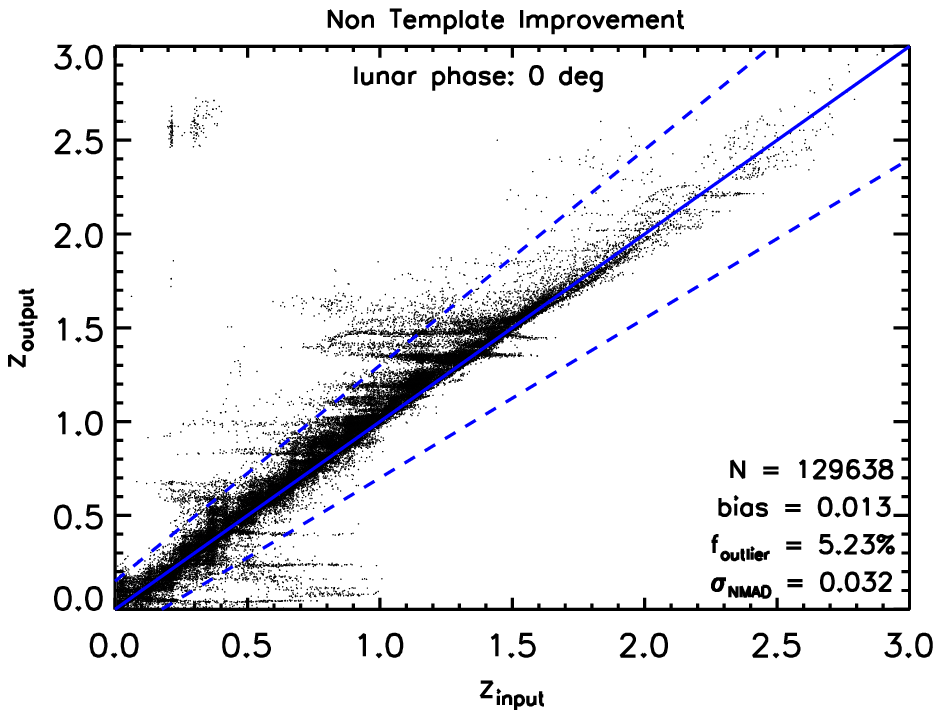}
   \includegraphics[width=0.49\columnwidth, angle=0]{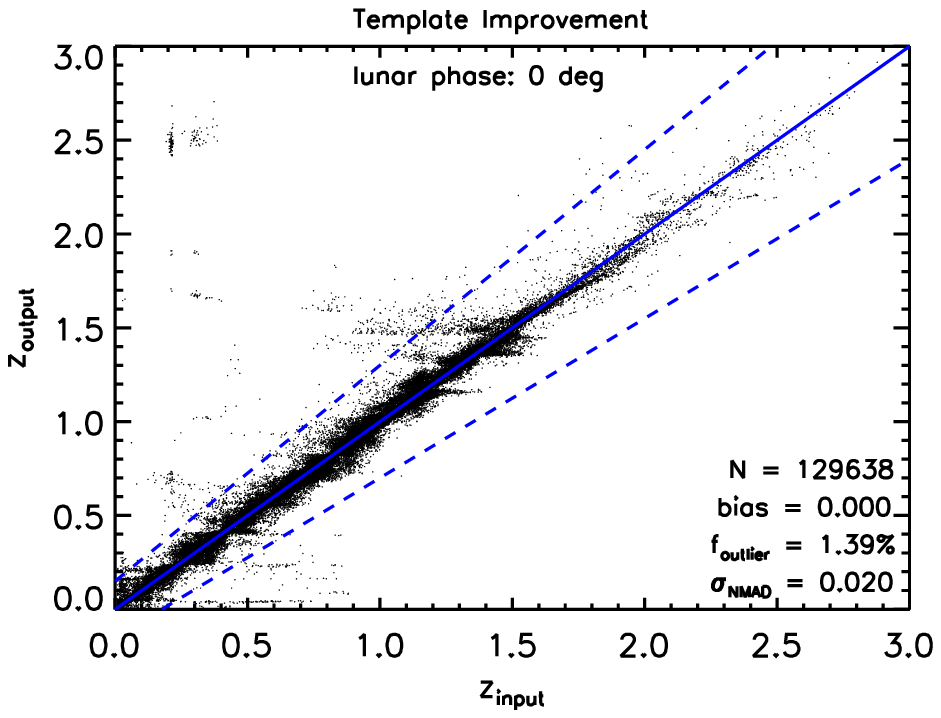}
   \includegraphics[width=0.49\columnwidth, angle=0]{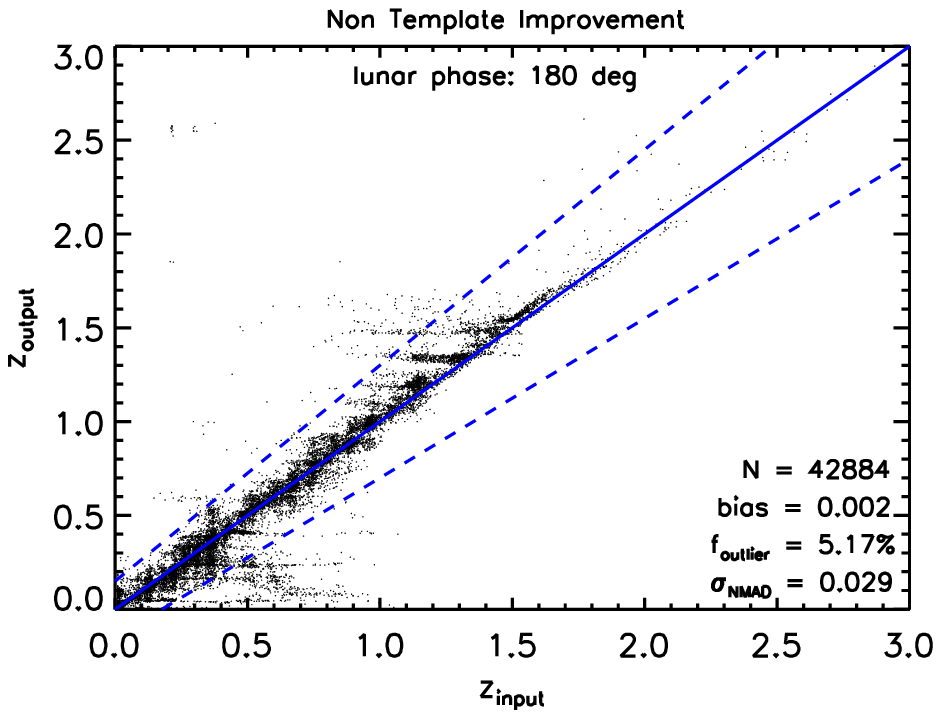}
   \includegraphics[width=0.49\columnwidth, angle=0]{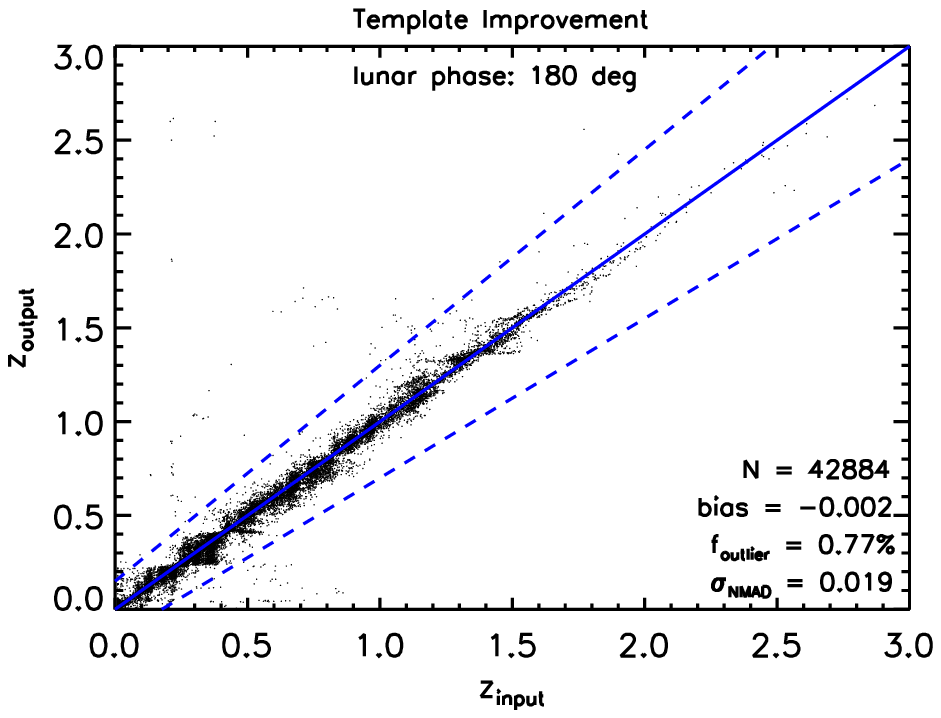}
   \caption{Similar to Figure~\ref{fig:zcom_shallow}, but for $z_{\rm phot}$ results in the deep mode with the addition of mock data from the {\it CSST}-$NUV$, {\it CSST}-$y$, {\it Euclid}-$Y_{\rm E}$, {\it Euclid}-$J_{\rm E}$, and {\it Euclid}-$H_{\rm E}$ bands.}
\label{fig:CSST_Euclid}
\end{figure}

Figure~\ref{fig:CSST_Euclid} shows the $z_{\rm phot}$ results in the deep mode with the addition of 5-band mock data from {\it CSST} and {\it Euclid}.
It is clear that the $z_{\rm phot}$ quality is significantly improved (cf. Figure~\ref{fig:zcom_deep}), because the 10-band mock photometry that well covers the wavelength from ultraviolet to near infrared is vital for both ZEBRA photometry-check mode and template-improvement mode.
In the non-template-improvement mode,
$f_{\rm outlier}$ and $\sigma_{\rm NMAD}$ are effectively reduced to $\sim5\%$ and $\sim$0.03, respectively;
lunar phase has little influence on $z_{\rm phot}$ results, mainly due to that mock {\it CSST} and {\it Euclid} data are almost unaffected by lunar phase.
In the template-improvement mode,
$f_{\rm outlier}$ and $\sigma_{\rm NMAD}$ are further reduced to $\sim1\%$ and $\sim$0.02, respectively; meanwhile, the bias is also better calibrated, being $\sim0.0$.

Fulfillment of many scientific goals is heavily dependent on $z_{\rm phot}$ accuracy, e.g.,
$z_{\rm phot}$ for future photometric weak lensing surveys need to at least achieve $\sigma_{\rm NMAD}<0.05$, with many relevant studies setting $\sigma_{\rm NMAD}\simeq0.02$ as a goal \citep[e.g.,][]{Zhan_2006,LSST_2009}, which is crucial to depict the redshift dependent growth of dark matter fluctuations, analyze weak lensing cosmic shears, and investigate the redshift dependent weak lensing signal behind clusters of galaxies under the constraints of the framework of the dark energy state equation \citep{Brimioulle_2008}.
As shown above, such requirements on $z_{\rm phot}$ accuracy can be met when the mock WFST,
{\it CSST} and {\it Euclid} are combined.

\section{Summary}\label{sect:sum}

\hspace{5mm}In this paper, we conduct a preliminary study that assesses $z_{\rm phot}$ qualtiy based on the mock WFST $ugriz$-band photometry in the shallow mode and deep mode.
We adopt the multiwavelength photometric catalog in the COSMOS/UltraVISTA field to generate mock WFST data, as it has deeper limiting magnitudes than WFST observations;
during this process, mock fluxes are computed through the convolution of galaxy SEDs with the 5 WFST filter transmission curves, with interstellar dust extinction and IGM absorption taken into account, and mock flux errors are evaluated through the consideration of instrumental parameters, sky background, and systematic errors.

We calculate $z_{\rm phot}$ using the ZEBRA code, which can generate new adaptive templates that better describe observed galaxy SEDs.
We find bias$\la0.006$, $\sigma_{\rm NMAD}\la0.03$, and $f_{\rm outlier}\la5\%$ in the shallow mode and
bias$\approx 0.005$, $\sigma_{\rm NMAD}\approx 0.06$, and $f_{\rm outlier}\approx 17\%$--$27\%$ in the deep mode, respectively, under various lunar phases;
lunar phase has limited influence on $z_{\rm phot}$ results, and the decrease of $z_{\rm phot}$ quality with dimming of lunar phase is primarily caused by sample-selection effect, i.e.,
the involvement of increasingly more fainter sources that have larger photometric uncertainties.

We compare our WFST $z_{\rm phot}$ results with that of some relevant works, finding general agreement between various results.
Given that the adaptive template fitting and machine learning methods
have their respective merits, it would be sensible to use all these methods jointly to further improve WFST $z_{\rm phot}$ quality in the future.

Finally, we compute $z_{\rm phot}$ by combining the mock WFST data with ultraviolet and near-infrared data from {\it CSST} and {\it Euclid}.
As expected, we find significant improvement in $z_{\rm phot}$ quality with $f_{\rm outlier}\approx 1\%$ and $\sigma_{\rm NMAD}\approx 0.02$, thanks to the full wavelength coverage from ultraviolet to near-infrared. Such high-quality $z_{\rm phot}$ can help fulfill many scientific goals that highly rely on $z_{\rm phot}$ accuracy.

\normalem
\begin{acknowledgements}

We thank the referee for a constructive report.
We thank Wen-tao Luo, Lu-lu Fan, and Ning Jiang for valuable discussions and comments.
This work is supported by
the National Key R\&D Program of China (2022YFF0503401),
the National Natural Science Foundation of China (12203047, 12025303, 11890693, and 12003031),
the science research grants from the China Manned Space Project with NO. CMS-CSST-2021-A06,
the Fundamental Research Funds for the Central Universities (WK3440000006 and WK2030000057),
the K.C. Wong Education Foundation, and
the Cyrus Chun Ying Tang Foundations.

\end{acknowledgements}

\bibliographystyle{raa}
\bibliography{bibtex}

\end{document}